\documentclass[twocolumn]{aastex63}
\usepackage[utf8]{inputenc}
\usepackage{soul}
\usepackage{float}
\usepackage{longtable}

\catcode`\@=11
\newcommand{\gsim}{${\mathrel{\mathpalette\@versim>}}$}
\newcommand{\lsim}{${\mathrel{\mathpalette\@versim<}}$}
\newcommand{\@versim}[2]{\lower 2.9truept \vbox{\baselineskip 0pt \lineskip
    0.5truept \ialign{$\m@th#1\hfil##\hfil$\crcr#2\crcr\sim\crcr}}}
\catcode`\@=12

\newcommand{\kms}{km s$^{-1}$}
\newcommand\HI{H\,{\sc i}}

\def\be{\begin{equation}}
\def\bea{\begin{eqnarray}}
\def\ee{\end{equation}}
\def\eea{\end{eqnarray}}

\begin{document}

\title{A Search for OH 18-cm Emission from Intermediate-Velocity Gas at High Galactic Latitudes}
\author[0000-0002-4772-9670]{Allison J. Smith}
\affiliation{Arecibo Observatory,
Arecibo, Puerto Rico 00612, USA}
\affiliation{University of Central Florida, 4000 Central Florida Blvd, Orlando, FL 32816, USA}

\author[0000-0002-1732-5990]{D. Anish Roshi}
\affiliation{Arecibo Observatory,
Arecibo, Puerto Rico 00612, USA}
\affiliation{University of Central Florida, 4000 Central Florida Blvd, Orlando, FL 32816, USA}

\begin{abstract}
    We present search results of 22 high latitude ($|b|$ $>$ 25$^{o}$) sightlines for OH 18-cm emission using the 305-m radio telescope at the Arecibo Observatory. These sightlines appear in neutral hydrogen emission at intermediate velocities $-90 \le V_{LSR} \le -20$ \kms\ and are predicted to have a sufficient molecular composition so as to be detectable in molecular emission. Such objects, known as Intermediate-Velocity Molecular Clouds (IVMCs), have historically been detected through $^{12}$CO emission. Recent studies indicate that IVMCs may be widespread in the Galaxy and have important implications for models of the interstellar medium and star formation. However, we report non-detections of OH emission toward the 22 sightlines and provide stringent upper limits on the OH column density. Using available HI and $A_v$ data in combination with existing state-of-the-art PDR models, we estimate H$_2$ column densities and find that they are more than an order of magnitude lower than the predicted values. We also find that the hydrogen volume density of these clouds is \lsim\ 25 cm$^{-3}$. In addition, we discuss the known IVMCs with previous $^{12}$CO detections in the context of the PDR models. Our analysis of these clouds indicates that the structure of molecular material in IVMCs is morphologically clumpy. These results motivate the need for future sensitive, on-the-fly searches (rather than targeted searches) for CO emission from IVMCs with of order $\sim$ 1\arcmin\ resolution. High angular resolution ($\sim$ 1\arcmin) HI and $A_v$ data will also be helpful to better constrain the structure and composition of IVMCs. 
    
\end{abstract}

\section{Introduction} \label{sec:int}
Molecular surveys of the Galactic plane during the 1980s and 1990s sought to answer a multitude of questions regarding the extent of molecular gas (especially with respect to how it extends into the lower halo) as well as the means via which atomic gas in our Galaxy transforms into molecular clouds and, ultimately, stars. 
During an early survey, \cite{des90} detected $^{12}$CO(1$-$0) emission from a high latitude sightline at an LSR velocity of $-$39 km s$^{-1}$, unusual for nearby gas comprising the high latitude extension of molecular gas in the plane. Later studies discovered still more molecular clouds at high latitude ($|b| \ge 25^{\circ}$) and intermediate velocities ($20 \le |v_{LSR}| \le 90$ km s$^{-1}$) \citep{des90, rea94, mag10}. Studies of their characteristics reveal that the intermediate velocity molecular clouds (IVMCs) are unique objects. Specifically, these clouds are significantly smaller (both spatially and by mass) than GMCs, less dense, and they also generally lack self-gravitating clumps and star formation. As such, they are superb laboratories for investigating the underlying mechanisms that dominate molecular cloud formation from atomic gas.

To date, the population of 
known IVMCs consists in its entirety of 11 distinct objects within five separate complexes. They are likely located at distances comparable to the atomic intermediate velocity clouds (IVCs), which are thought to lie at roughly 1-2 kpc {\citep{wak01}}. The IVMCs are distinct, however, as they have a sufficient molecular composition so as to be detected in CO(1$-$0) emission rather than solely in H$_2$ absorption such as the IVCs studied with FUSE \citep[see, e.g.,][]{ric03}. While the origins of all thick-disk and halo clouds (including high velocity clouds) are convoluted yet important for understanding the evolution of our Galaxy \citep{put12}, current evidence indicates that atomic IVCs probably originate in the lower halo as material from the Galactic fountain \footnote{In the Galactic fountain model material is deposited into the lower halo from energetic processes in the disk where it later cools and rains down to the Galactic plane \citep{bre80}.}. A good means of assessing whether IVMCs also originate in the fountain is to measure their metallicities to determine whether they match the composition of the disk. One such study has been done for IVMC G135$+$51.3 and indicated the cloud metallicity is subsolar  \citep[about 0.43 dex;]{her13}, implying a fountain origin may be less likely, and further illuminating the importance of understanding the characteristics and formation mechanisms of these objects. The question of their origins, the insight they can provide into the physics of the atomic to molecular transition, and their potential influence on the Galactic star formation cycle, especially if they are a widespread phenomenon, are all important motivations for studying these clouds. 

The 353-GHz all sky emission observed by the Planck spacecraft presented an improved chance of addressing the question of the distribution and prevalence of IVMCs. This emission is dominated by thermal emission from Galactic dust grains. Using compiled HI survey data from EBHIS \citep{win16} and Parkes \citep{bar01} and the 353-GHz map from Planck, \cite{roh16} created a map of candidate IVMCs by employing the HI-FIR correlation \citep{bou96}. They produced a map of 239 candidate IVMCs in the northern and southern Galactic hemispheres, all of which are predicted to have H$_2$ column densities N(H$_2$) $>$ 10$^{19}$ cm$^{-2}$. 

In order to probe low column density molecular gas, a CO or OH absorption study would in principle be a useful approach. However, considering the dearth of bright background sources coincident with the candidate IVMCs, the combination of surface brightness sensitivity and high angular resolution of the 305-m Arecibo legacy telescope, and the success with which we previously observed OH 18-cm emission from an IVMC \citep{smi18}, we elected to search for OH 18-cm emission toward a subset of candidates listed in the \cite{roh16} catalogue.

We describe in Section~\ref{obs} the justification for using OH as a tracer, the target selection process, the observing setup and data reduction method used in the paper. We present our results in Section~\ref{res} and the interpretation of the results in Section~\ref{disc}. In Section~\ref{sum}, we provide a summary of the observations and the main results and discuss implications for future observational investigations of IVMCs. 

\section{IVMC Tracers, Target selection, Observations and data processing}
\label{obs}

We observed 22 sightlines within 11 candidate IVMCs listed in \cite{roh16} (see Table~\ref{tab1}) for OH 18-cm emission with the 305-m radio telescope at the Arecibo Observatory in Arecibo, Puerto Rico. The project (proj. code A3344) was initially granted 63 hours of observing time, and we carried out the observations primarily between 31 July 2019 and March 2020. In this section, we discuss the rationale behind selecting the OH 18-cm lines for the study, followed by the target selection process, the observational setup and data processing method.

\centerwidetable
\startlongtable
\begin{deluxetable*}{lrrrrrrrrrr}
\tablecaption{Observational Parameters of Target Sightlines \label{tab1}}
\tablewidth{0pt} 
\tablehead{ 
\colhead{Source} & \colhead{RA} & \colhead{DEC}& \colhead{\it l} &\colhead{\it b} &\colhead{V$_{LSR}$\tablenotemark{a}} & \colhead{T$_B$\tablenotemark{a}} & \colhead{$\Delta V$\tablenotemark{a}} & \colhead{Int. Time} & \colhead{ Upper limit}\tablenotemark{b}\\
 & \colhead{(hh mm ss)}  &\colhead{($^\circ$ $^\prime$ $^{\prime\prime}$)} & \colhead{($^\circ$)} & \colhead{($^\circ$)} & \colhead{(\kms)} & \colhead{(K)} & \colhead{(\kms)} & \colhead{(minutes)} & \colhead{(mK)}}
 
\startdata
\hline
\multicolumn{11}{c}{Observed Positions}\\
\hline
IVMC01      & 22 45 49.3  & +16 43 41.4  & 84.5 & -36.6 & -50.75$\pm$0.05 &  1.56$\pm$0.03 &  6.83$\pm$0.17 &  20 &  8.3 \\
            &             &              &      &       & -45.00$\pm$0.32 &  1.50$\pm$0.02 & 25.49$\pm$0.55 &     &      \\
IVMC01N     & 22 45 49.3  & +16 49 41.4  & 84.6 & -36.5 & -51.78$\pm$0.01 &  5.36$\pm$0.03 &  5.80$\pm$0.04 &  10 &  9.9 \\
            &             &              &      &       & -47.50$\pm$0.12 &  2.34$\pm$0.02 & 22.92$\pm$0.25 &     &      \\
IVMC01E     & 22 45 25.3  & +16 43 41.4  & 84.4 & -36.5 & -49.40$\pm$0.02 &  4.35$\pm$0.05 &  3.20$\pm$0.05 &  20 & 10.6 \\
            &             &              &      &       & -48.65$\pm$0.14 &  2.06$\pm$0.03 & 20.43$\pm$0.35 &     &      \\
IVMC01W     & 22 46 13.3  & +16 43 41.4  & 84.6 & -36.7 & -51.02$\pm$0.03 &  3.89$\pm$0.07 &  6.70$\pm$0.12 &  20 & 15.7 \\
            &             &              &      &       & -46.70$\pm$0.26 &  3.03$\pm$0.06 & 22.87$\pm$0.52 &     &      \\
            &             &              &      &       & -19.90$\pm$0.24 &  2.85$\pm$0.03 & 18.17$\pm$0.78 &     &      \\
IVMC10      & 10 17 27.1  & +28 45 26.1  & 201.0 & 56.1 & -50.98$\pm$0.01 &  8.01$\pm$0.05 &  3.97$\pm$0.03 & 590 &  3.6 \\
            &             &              &      &       & -45.55$\pm$0.13 &  1.92$\pm$0.02 & 26.73$\pm$0.30 &     &      \\
IVMC15      & 09 32 35.3  & +21 23 26.8  & 208.9 & 44.6 & -37.91$\pm$0.07 &  1.40$\pm$0.02 &  8.55$\pm$0.13 &  65 &  2.8 \\
            &             &              &      &       & -26.63$\pm$0.10 &  1.02$\pm$0.02 & 10.89$\pm$0.35 &     &      \\
IVMC26      & 09 39 43.7  & +14 07 40.6  & 219.2 & 43.6 & -35.55$\pm$0.05 &  4.04$\pm$0.13 &  7.30$\pm$0.15 & 190 &  4.6 \\
            &             &              &      &       & -25.37$\pm$0.04 &  4.03$\pm$0.07 &  5.13$\pm$0.10 &     &      \\
            &             &              &      &       & -28.88$\pm$0.50 &  2.00$\pm$0.00 & 15.21$\pm$0.66 &     &      \\
IVMC27      & 10 12 50.5  & +16 35 31.0  & 220.6 & 51.9 & -40.97$\pm$0.78 &  1.04$\pm$0.02 & 29.44$\pm$1.27 & 200 &  4.2 \\
            &             &              &      &       & -32.70$\pm$0.03 &  3.23$\pm$0.05 &  4.93$\pm$0.09 &     &      \\
            &             &              &      &       & -20.42$\pm$0.41 &  1.00$\pm$0.00 & 14.37$\pm$0.83 &     &      \\
IVMC28      & 12 12 50.7  & +23 51 19.4  & 231.4 & 80.7 & -44.48$\pm$0.02 &  1.91$\pm$0.02 &  5.02$\pm$0.07 &  95 &  4.4 \\
            &             &              &      &       & -46.41$\pm$0.24 &  1.52$\pm$0.02 & 23.59$\pm$0.36 &     &      \\
IVMC28N     & 12 12 50.7  & +23 57 19.4  & 230.8 & 80.7 & -45.80$\pm$0.04 &  1.50$\pm$0.00 &  6.69$\pm$0.11 &  70 &  5.9 \\
            &             &              &      &       & -50.99$\pm$0.26 &  1.05$\pm$0.02 & 26.98$\pm$0.42 &     &      \\
IVMC28S     & 12 12 50.7  & +23 45 19.4  & 231.9 & 80.7 & -44.45$\pm$0.09 &  1.50$\pm$0.00 &  6.87$\pm$0.22 &  75 &  4.4 \\
            &             &              &      &       & -51.77$\pm$0.22 &  0.78$\pm$0.04 & 10.79$\pm$0.47 &     &      \\
            &             &              &      &       & -28.19$\pm$0.71 &  1.27$\pm$0.02 & 49.07$\pm$0.69 &     &      \\
IVMC28E     & 12 12 26.7  & +23 51 19.4  & 231.2 & 80.6 & -46.07$\pm$0.06 &  1.25$\pm$0.04 &  9.01$\pm$0.22 &  30 &  6.4 \\
            &             &              &      &       & -50.74$\pm$0.36 &  0.97$\pm$0.04 & 21.03$\pm$0.39 &     &      \\
IVMC28W     & 12 13 14.7  & +23 51 19.4  & 231.6 & 80.8 & -44.09$\pm$0.05 &  2.01$\pm$0.04 &  5.92$\pm$0.15 &  45 &  6.0 \\
            &             &              &      &       & -41.82$\pm$0.48 &  1.33$\pm$0.03 & 31.69$\pm$1.05 &     &      \\
            &             &              &      &       & -27.09$\pm$0.24 &  0.54$\pm$0.05 &  7.31$\pm$0.82 &     &      \\
IVMC29      & 11 52 06.0  & +17 04 46.7  & 246.1 & 73.0 & -30.45$\pm$0.09 &  2.49$\pm$0.01 & 27.24$\pm$0.18 & 270 &  5.7 \\
            &             &              &      &       & -22.71$\pm$0.01 &  4.45$\pm$0.03 &  3.95$\pm$0.03 &     &      \\
IVMC30      & 12 01 57.9  & +17 06 42.2  & 251.6 & 74.8 & -29.59$\pm$0.03 &  4.14$\pm$0.05 &  4.56$\pm$0.07 & 135 &  5.5 \\
            &             &              &      &       & -18.95$\pm$0.38 &  1.27$\pm$0.02 & 43.47$\pm$0.70 &     &      \\
IVMC36      & 12 24 29.6  & +20 02 01.8  & 260.5 & 80.6 & -29.23$\pm$0.05 &  2.70$\pm$0.00 & 14.72$\pm$0.11 & 170 &  6.3 \\
            &             &              &      &       & -27.79$\pm$0.04 &  1.77$\pm$0.03 &  3.49$\pm$0.08 &     &      \\
IVMC37      & 12 18 52.7  & +17 46 06.7  & 262.5 & 78.0 & -30.26$\pm$0.05 &  3.00$\pm$0.00 &  7.39$\pm$0.13 & 115 &  4.4 \\
IVMC42      & 12 37 18.3  & +20 28 12.9  & 276.3 & 82.6 & -28.98$\pm$0.02 &  4.41$\pm$0.03 &  6.02$\pm$0.05 & 115 &  5.0 \\
            &             &              &      &       & -19.95$\pm$0.45 &  0.85$\pm$0.01 & 43.37$\pm$0.85 &     &      \\
IVMC42N     & 12 37 18.3  & +20 34 12.9  & 275.9 & 82.7 & -28.91$\pm$0.02 &  3.83$\pm$0.04 &  5.71$\pm$0.07 &  80 &  7.5 \\
            &             &              &      &       & -28.22$\pm$0.14 &  1.22$\pm$0.03 & 24.70$\pm$0.49 &     &      \\
IVMC42S     & 12 37 18.3  & +20 22 12.9  & 276.6 & 82.5 & -28.67$\pm$0.02 &  3.16$\pm$0.04 &  3.48$\pm$0.05 &  70 &  7.5 \\
            &             &              &      &       & -28.72$\pm$0.09 &  1.41$\pm$0.03 & 17.84$\pm$0.28 &     &      \\
IVMC42E     & 12 36 54.3  & +20 28 12.9  & 275.6 & 82.6 & -28.74$\pm$0.02 &  4.69$\pm$0.04 &  4.87$\pm$0.06 &  25 &  6.3 \\
            &             &              &      &       & -26.89$\pm$0.17 &  1.21$\pm$0.03 & 22.41$\pm$0.51 &     &      \\
IVMC42W     & 12 37 42.3  & +20 28 12.9  & 276.9 & 82.6 & -28.91$\pm$0.02 &  2.74$\pm$0.04 &  4.21$\pm$0.07 &  55 &  5.8 \\
            &             &              &      &       & -28.87$\pm$0.13 &  1.60$\pm$0.03 & 18.53$\pm$0.36 &     &      \\
\hline
\multicolumn{11}{c}{Summed Positions}\\
\hline
IVMC01T     & 22 45 49.3  & +16 43 41.4  & 84.5 & -36.6 & -50.75$\pm$0.03 &  2.45$\pm$0.02 &  7.48$\pm$0.09 &  70 &  4.9 \\
            &             &              &      &       & -39.38$\pm$0.37 &  1.19$\pm$0.01 & 35.83$\pm$0.66 &     &      \\
IVMC28T     & 12 12 50.7  & +23 51 19.4  & 231.4 & 80.7 & -44.15$\pm$0.05 &  1.50$\pm$0.00 &  7.24$\pm$0.13 & 315 &  3.7 \\
            &             &              &      &       & -44.90$\pm$0.35 &  1.22$\pm$0.02 & 28.50$\pm$0.58 &     &      \\
IVMC42T     & 12 37 18.3  & +20 28 12.9  & 276.3 & 82.6 & -28.86$\pm$0.02 &  3.39$\pm$0.03 &  4.36$\pm$0.05 & 345 &  4.7 \\
            &             &              &      &       & -28.77$\pm$0.11 &  1.58$\pm$0.03 & 18.10$\pm$0.32 &     &      &\\
\enddata
\tablenotetext{a}{\HI\ 21-cm line parameters of the components in the velocity range $-$20 to $-$90 \kms. Parameters with no errors are those that were fixed while fitting Gaussian components to the spectrum.}
\tablenotetext{b}{ 3$\sigma$ value of the noise fluctuations in the 1667 MHz OH spectrum obtained after smoothing the final spectrum by the smallest width of the identified IVMC components listed in the table toward each observed position. The RMS values were obtained from the data in the velocity range $-$20 to $-$90 \kms\ after subtracting a 3$^{rd}$ order polynomial.} 
\end{deluxetable*}
\subsection{Tracers for IVMCs}
\label{tracers}

Traditionally, the $^{12}$CO(1$-$0) transition at 115 GHz has been the workhorse tracer for detecting and characterizing high latitude molecular clouds in the ISM \citep{mag85, mag96, mag00, mag10}. These studies ultimately uncovered over 300 of these objects. However, high-latitude molecular clouds are extended sources, sometimes extending a degree or more in angular size, and they are generally closer than the known IVMC population, the latter being much smaller in angular size (half a degree or so) due to their typical distances ($\sim 1-2$ kpc). Therefore, with respect to detecting IVMCs, the large-scale surveys were likely limited, both in required sky coverage (as these surveys were typically untargeted with 1$^\circ$ sampling) and in sensitivity, as they typically consisted of 5 minute on/off observing patterns that reached noise levels of order tens of mK \citep[see, e.g.,][]{mag96}.

Targeted CO surveys for IVMCs have had a somewhat higher detection rate, as \cite{mag10} detected intermediate velocity CO in 11 out of 77 sightlines using color excess as a guide. However, among the few IVMCs for which CO observations exist (henceforth referred to as ``known IVMCs''), the CO structure appears to be somewhat clumpy within the cloud, with detections typically clustered together. Hence, unless telescope availability allows for checking multiple sightlines in each cloud, a deep, large-scale search for IVMCs using CO is often impractical.

Observing the OH 18-cm ground state transition is generally also a reliable means of tracing molecular gas. Additionally, it has been observed with the Arecibo telescope in emission in multiple locations in previously identified IVMC G211$+$63 \citep{smi18}, in which the OH structures stretch 10$\arcmin$ or more across the cloud. One important caveat, however, is that the excitation temperature of the OH molecule in these environments within the diffuse ISM is frequently quite close to the background CMB temperature \citep{bus19,tan20} in which case one may not observe the line in emission. In this case, an absorption study is required to detect the gas. Unfortunately, choices for bright background sources for our targets are sparse, but an OH absorption study is most likely to be successful when viable background sources are available.

\begin{deluxetable*}{lcrrrcccccc}[htbp]
\tabletypesize{\scriptsize} 
\tablecaption{Derived Physical Parameters of Selected Targets
\label{tab2}}
\tablewidth{0pt} 
\tablehead{\colhead{(1)} & \colhead{(2)} & \colhead{(3)} &\colhead{(4)}  &\colhead{(5)} &\colhead{(6)} &\colhead{(7)} &\colhead{(8)} &\colhead{(9)} & \colhead{(10)} & \colhead{(11)} \\
\colhead{Source} & \colhead{V$_{LSR}$} & \colhead{$A_v$} &\colhead{IV$A_v$}  &\colhead{N(HI$_{IV}$)} &\colhead{N(HI$_{IV}$)} &\colhead{N(H$_2$)} &\colhead{N(H$_2$)} &\colhead{N(OH)$_{10}$} & \colhead{N(OH)$_{5}$} & \colhead{N(H$_2$)}\\ 
\tabletypesize{\small}& \colhead{(km s$^{-1}$)} & \colhead{(mag)} & \colhead{(mag)} & \colhead{(10$^{20}$cm$^{-2}$)} & \colhead{(10$^{20}$cm$^{-2}$)} & \colhead{(10$^{20}$cm$^{-2}$)} & \colhead{(10$^{20}$cm$^{-2}$)} &\colhead{(10$^{13}$cm$^{-2}$)} & \colhead{(10$^{13}$cm$^{-2}$)} & \colhead{(10$^{20}$cm$^{-2}$)} }
\startdata
\hline
\multicolumn{11}{c}{Candidate IVMCs}\\
\hline
IVMC01   & -53.10 &  0.22 &  0.03$\pm$0.04 &  1.36$\pm$0.06 &  0.95$\pm$0.02 &  2.61$\pm$0.18 & -0.21$\pm$0.48 &   3.3 &   6.1 &   6.1     \\
IVMC10  & -52.30 &  0.16 &  0.14$\pm$0.03 &  2.19$\pm$0.05 &  1.62$\pm$0.02 &  0.38$\pm$0.12 &  0.73$\pm$0.35 &   0.8 &   1.5 &   1.5     \\
IVMC15  & -39.50 &  0.09 &  0.00$\pm$0.01 &  0.51$\pm$0.02 &  0.45$\pm$0.01 &  5.47$\pm$0.54 & -0.24$\pm$0.20 &   1.3 &   2.4 &   2.4     \\
IVMC26  & -38.60 &  0.19 &  0.11$\pm$0.03 &  2.21$\pm$0.03 &  1.57$\pm$0.04 &  0.71$\pm$0.16 &  0.40$\pm$0.41 &   1.3 &   2.4 &   2.4     \\
IVMC27  & -34.80 &  0.12 &  0.06$\pm$0.03 &  2.22$\pm$0.17 &  1.19$\pm$0.03 & 42.10$\pm$14.40 & -0.33$\pm$0.37 &   1.2 &   2.2 &   2.2     \\
IVMC28  & -20.20 &  0.12 &  0.04$\pm$0.02 &  1.06$\pm$0.06 &  0.89$\pm$0.01 &  0.40$\pm$0.08 &  0.07$\pm$0.27 &   1.3 &   2.4 &   2.4     \\
IVMC29  & -26.00 &  0.16 &  0.11$\pm$0.03 &  2.19$\pm$0.03 &  1.67$\pm$0.01 & 11.95$\pm$0.65 &  0.45$\pm$0.35 &   1.3 &   2.4 &   2.4     \\
IVMC30  & -30.70 &  0.12 &  0.09$\pm$0.02 &  1.81$\pm$0.05 &  1.44$\pm$0.02 &  9.96$\pm$0.89 &  0.26$\pm$0.26 &   1.4 &   2.6 &   2.6     \\
IVMC36  & -29.50 &  0.09 &  0.03$\pm$0.01 &  1.39$\pm$0.04 &  0.90$\pm$0.01 &  1.26$\pm$0.50 & -0.23$\pm$0.20 &   1.3 &   2.3 &   2.3     \\
IVMC37  & -31.50 &  0.12 &  0.04$\pm$0.02 &  0.53$\pm$0.03 &  0.43$\pm$0.01 &  8.60$\pm$2.84 &  0.28$\pm$0.26 &   1.9 &   3.5 &   3.5     \\
IVMC42  & -30.10 &  0.09 &  0.03$\pm$0.01 &  1.46$\pm$0.04 &  1.24$\pm$0.02 &  0.78$\pm$0.19 & -0.26$\pm$0.19 &   1.8 &   3.3 &   3.3     \\
\hline
\multicolumn{11}{c}{Summed Positions}\\
\hline
IVMC01T &    &  0.22 &    &    &  1.19$\pm$0.02 &    &    &   2.1 &   3.8 &   3.8     \\
IVMC28T &    &  0.12 &    &    &  0.89$\pm$0.02 &    &    &   1.5 &   2.8 &   2.8     \\
IVMC42T &    &  0.09 &    &    &  0.85$\pm$0.01 &    &    &   1.2 &   2.2 &   2.2    
\enddata
\tablenotetext{}{Columns:(1)Target name. In addition to the 11 sightlines from the \cite{roh16} candidate list, three summed positions are included from the five-point cross patterns (indicated with ``T"); (2)LSR velocity for the IVMC complexes provided by \cite{roh16};(3) 
$A_v$ obtained from \cite{sch98}. The angular resolution of their $A_v$ image is 6.1\arcmin\ and the uncertainty in $A_v$ is up to 16\%;(4)$A_v$ of the IVMC estimated in this work (see Section~\ref{res}); (5)HI column density obtained from GALFA data after smoothing the data to $\sim$6\arcmin\ resolution; (6)HI column density obtained from the IVMC components listed in Table~\ref{tab1} (angular resolution $\sim$ 3\arcmin);(7)H$_2$ column density from \cite{roh16};(8) H$_2$ column density estimated using Eq.~\ref{NH2}. We used values tabulated in columns (4) and (5) for $A_v$ and N(HI) respectively, which have the same angular resolution, to get the H$_2$ column density;(9)OH column density estimated using the upper limits listed in Table~\ref{tab1} for an excitation temperature of 10 K. We assumed that the OH emission has the smallest IVMC HI line width in each target to get the column density; (10)OH column density estimated using the upper limits listed in Table~\ref{tab1} for an excitation temperature of 5 K. We assumed that the OH emission has the smallest IVMC HI line width in each target to get the column density; (11) Upper limit on H$_2$ column density obtained from values tabulated in column (10) using an $X_{OH} \sim 10^{-7}$ (see Section~\ref{modelcompare}).}
\end{deluxetable*}

\subsection{Target Selection} \label{subsection:tar}

In order to systematically observe the candidate IVMCs in \cite{roh16} for OH emission, we elected to observe as many sightlines as possible while reaching a specific noise level threshold. Previous OH detections in G211$+$63 dictated that we may expect lines as weak as 10 mK. Hence, we decided to aim for \lsim\ 2 mK RMS noise levels at each position. There were 42 candidate IVMCs within the Arecibo telescope declination range. We focused on 11 initial targets by prioritizing sightlines with either high N(H$_2$) as predicted by \cite{roh16} or high visual extinction ($A_v$) as given by the \cite{sch98} dust maps. The selected candidates have predicted N(H$_2$) spanning roughly two orders of magnitude from $\sim4\times10^{19}$ to $4 \times 10^{21}$ cm$^{-2}$ (see Table~\ref{tab2}). \cite{smi18} detected OH emission in regions with $A_v$ $\sim$ 0.25 mag. Our sample for the present study, however, was constrained to lower $A_v$ sources by the declination range of the telescope, and the values for the selected candidates span from 0.09 to 0.22 mag (see Table~\ref{tab2}). Hence, this study probes an $A_v$ parameter space up to the previous threshold.

\begin{figure*}[htbp]
    \centering
     \epsscale{.9}\plottwo{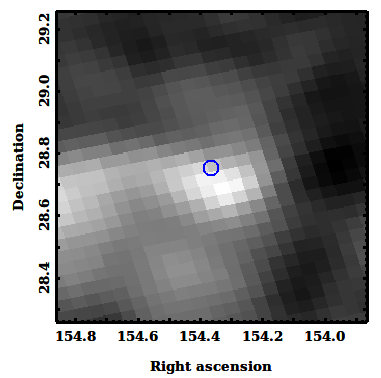}{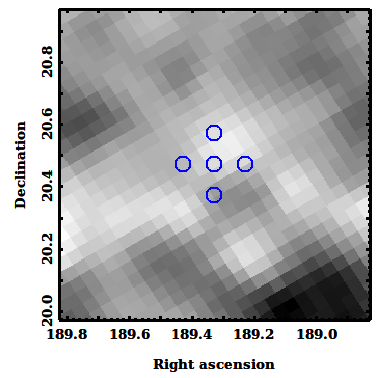}
    \caption{\cite{sch98} dust maps of IVMC10 (left) and IVMC42 (right) in our study. Both images are 60$\arcmin$ across. Light gray indicates higher E(B$-$V), and dark gray indicates lower. The conversion factor used in this study to obtain A$_v$ is 3.1. The E(B-V) values in these images vary from 0.02-0.05 for IVMC10, and from 0.03-0.035 for IVMC42. For IVMC10 we observed a single position at the coordinates listed in \cite{roh16} which were highly consistent with appears to be peak dust and HI emission (based on GALFA data). For IVMC42, we observed a 5-point cross pattern centered on the \cite{roh16} coordinates since the color excess changed so little across the entire inner portion covered by the 16$\arcmin$ HI4PI beam. The cross pattern spans the full HI4PI beam. Each circle in the cross pattern represents the central, north, south, east and west positions and as covered by the 3$\arcmin$ Arecibo beam.}
    \label{fig1}
\end{figure*}

The angular resolution of the \cite{roh16} study (16$\arcmin$) was determined by the resolution of the HI4PI survey data \citep{hi4pi}. Given that the Arecibo telescope beam was $\sim$ 3$\arcmin$ at 1.6 GHz we examined the morphology of each cloud in HI emission using GALFA-HI data cubes \citep{pee18} and the $A_v$ values from the \cite{sch98} dust maps (angular resolution $\sim$ 6$\arcmin$) to carefully select pointing locations within each cloud. In every case, the HI data cubes showed bright, extended emission at the intermediate velocities, and the $A_v$ varies little within the cloud. There are a variety of morphologies, as some appear more extended, while others have a clumpy structure in dust and HI emission, the former extending over several arcminutes before dropping off quickly by as much as a factor of five. As telescope time allowed, we attempted 5-point cross patterns in right ascension and declination. on three sources whose morphology was more extended since there was no obvious central location at which to point. The cross pattern was selected to partially cover the full HI4PI beam, ensuring that multiple positions were checked within the 16$\arcmin$ beam and also providing the opportunity to combine the data from all 5 positions to achieve a more sensitive detection limit. Examples of the different morphologies and observing patterns are shown in Fig.~\ref{fig1}.

These 11 initial targets, the individual five-point cross positions (designated as East, West, North and South), and the summed positions (Total, consisting of the cross and center positions) are listed alongside their observed parameters in Table~\ref{tab1}. We thus observed a total of 22 separate sightlines for OH emission. The source name, J2000 right ascension and declination, Galactic longitude and latitude, intermediate-velocity HI line parameters (Local standard of rest velocity, peak brightness temperature, and full width at half maximum line width) of each component, and finally the integration time and 3$\sigma$ noise levels for our OH observations are listed in Table~\ref{tab1}.



\subsection{Observational Setup}

 We used the ``L-band Wide'' receiver and configured the Wideband Arecibo Pulsar Processors (WAPPs) to observe simultaneously both polarizations of the OH 18-cm main and satellite lines at 1665, 1667, 1612, and 1720 MHz, as well as the 21-cm line at 1420 MHz. We used a bandwidth of 1.5625 MHz with 9-level sampling and 2048 channels per WAPP subcorrelator. Some of the molecular lines from high-latitude regions can be quite narrow (e.g., 0.7 \kms), and this setup yields a resolution of better than 0.14 \kms\ per channel. The expected line widths are much narrower than typical spectral bandpass variations of the system (typically tens of \kms), so we used an observing pattern consisting of only ON-mode for these observations. Each ON scan was 300 seconds long followed by a one second integration with a calibrated noise source and one second integration without the noise source injection. Multiple such scans were obtained toward each target. The total on-source observing times for different target sightlines are listed in Table~\ref{tab1}.
 
 \subsection{Data Reduction} 
\label{dat}

\begin{figure*}[ht!]
    \centering
    \includegraphics[width=6in]{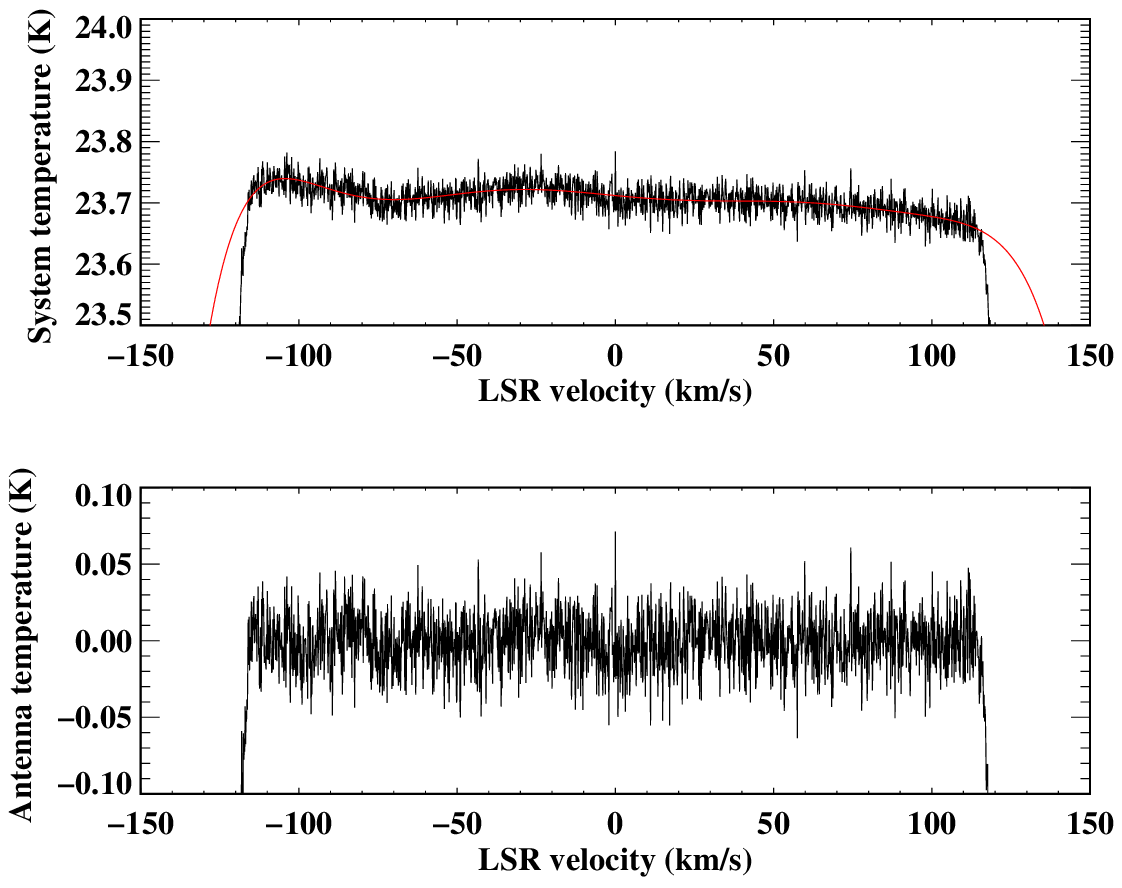}
    \caption{The average OH spectrum toward the source IVMC28N and the 8th order polynomial model of the bandpass are shown on the top panel. The bandpass corrected spectrum is shown in the bottom panel.}
    \label{fig2}
\end{figure*}

We processed these data using existing\footnote{\url{http://www.naic.edu/~phil/software/software.html##idldoc}} and new routines developed with the Interactive Data Language (IDL) package. The data processing begins by averaging the 300 
one-second integrations in each ON scan. The data from each polarization are separately averaged, and the averaged spectrum is then calibrated in units of antenna temperature using the noise injection scans. Upon visual inspection of each calibrated spectrum, we averaged all the ON scans obtained on a day toward a target and then averaged the spectra from the two orthogonal polarizations. We modeled the bandpass by fitting a higher (often 8$^{\rm{th}}$) order polynomial to the manually identified line-free channels of the calibrated spectrum. The line-free channels were chosen using the predicted velocities in the \cite{roh16} study which were also generally consistent with the HI velocities we observed in our data. We hence selected line-free channels in visually flat portions of the spectrum that were at least 20 \kms away from the expected velocity. We performed the bandpass correction dividing the spectrum by the polynomial fit; we then subtracted unity to make the mean of the noise fluctuations near zero. We show an example of a bandpass model and the corrected spectrum toward one of the targets in Fig.~\ref{fig2}. Although the Arecibo Observatory is in the Puerto Rico Coordination Zone, we did not request coordination for the particularly vulnerable 1612 MHz line since we saw only non-detections in the main lines in the initial data. The only other significant RFI was instrumental in nature in the 1720 MHz band. We had no RFI issues in the main lines. Our analysis focuses on the 1667 MHz line data since we also did not detect any emission in the 1665 MHz line.

During the observations, we did not apply the online Doppler correction, as the shift in velocity was within $\pm$ 0.02 \kms. To align the spectral line in the LSR (Local Standard of Rest) frame for different days observations, we used an FFT (Fast Fourier Transform) interpolation method to re-sample the bandpass corrected spectrum before averaging them together to obtain the final spectrum toward a position. The HI line was detected immediately in every case. Gaussian modeling of the HI line (see Appendix~\ref{A}) was used to get the parameters of the intermediate-velocity HI lines, which are listed in Table~\ref{tab1}. The final OH spectrum was smoothed to a velocity resolution equal to the smallest intermediate-velocity HI line width toward each position. The upper limits, listed in Table~\ref{tab1}, were the 3$\sigma$ values computed from the data points over the velocity range -90 to -20 \kms\ after subtracting a 3$^{rd}$ order polynomial from the smoothed spectrum.  We use these values in Section~\ref{res} to obtain the upper limits on OH column density, N(OH), for each position.

\section{Results} 
\label{res}

\begin{deluxetable*}{lccccccccc}
\tabletypesize{\scriptsize} 
\tablecaption{Observed and estimated parameters of known IVMCs\label{tab3}}
\tablewidth{0pt} 
\tablehead{\colhead{(1)} &\colhead{(2)} &\colhead{(3)}  &\colhead{(4)} &\colhead{(5)} &\colhead{(6)} &\colhead{(7)} &\colhead{(8)} &\colhead{(9)} &\colhead{(10)}\\ 
\colhead{Source} &\colhead{V$_{LSR(CO)}$} &\colhead{W$_{CO}$}  &\colhead{$\int$ T$_B^{tot HI}$d$\nu$} &\colhead{$\int$ T$_B^{LV HI}$ d$\nu$} &\colhead{$\int$ T$_B^{IV HI}$ d$\nu$} &\colhead{N(HI$_{IV}$)} &\colhead{$A_v$} &\colhead{IV $A_v$} &\colhead{Ref}\\ 
&\colhead{(km s$^{-1}$)} &\colhead{(K km s$^{-1}$)} &\colhead{(K km s$^{-1}$)} &\colhead{(K km s$^{-1}$)} &\colhead{(K km s$^{-1}$)} &\colhead{(10$^{20}$ cm$^{-2}$)} &\colhead{(mag.)} &\colhead{(mag.)} &}
\startdata
G210.8$+$63.1 & -38.7 &  0.8$\pm$0.3 & 199$\pm$1 &    31$\pm$1 &   167$\pm$1 &   3.05$\pm$0.02 &   0.28 &   0.26$\pm$0.04 &   a\\
G135.2$+$54.5 & -45.2 &  1.0$\pm$0.2 & 176$\pm$4 &    17$\pm$2 &   151$\pm$3 &   2.76$\pm$0.05 &   0.22 &   0.21$\pm$0.04 &   b\\
G135.5$+$51.3\tablenotemark{1} & -48.0 &  0.2$\pm$0.1 & 184$\pm$4 &    14$\pm$2 &   166$\pm$2 &   3.03$\pm$0.04 &   0.19 &   0.18$\pm$0.03 &   b\\
G283.8$+$54.9 & -33.8 &  0.3$\pm$0.1 & 254$\pm$2 &    79$\pm$1 &   173$\pm$1 &   3.16$\pm$0.02 &   0.37 &   0.32$\pm$0.06 &   c\\
G288.4$+$53.2 & -24.4 &  0.9$\pm$0.1 & 274$\pm$2 &   130$\pm$1 &   136$\pm$1 &   2.48$\pm$0.02 &   0.40 &   0.31$\pm$0.06 &   c\\
G295.0$+$57.1 & -21.6 &  0.4$\pm$0.1 & 203$\pm$2 &   144$\pm$1 &    56$\pm$1 &   1.03$\pm$0.02 &   0.25 &   0.15$\pm$0.04 &   c\\
G91.3$+$38.0  & -22.7 &  3.4$\pm$0.3 & 128$\pm$1 &    62$\pm$1 &    65$\pm$1 &   1.19$\pm$0.01 &   0.19 &   0.15$\pm$0.03 &   d\\
G89.8$+$38.7  & -24.1 &  1.4$\pm$0.3 & 154$\pm$2 &    71$\pm$1 &    81$\pm$1 &   1.48$\pm$0.02 &   0.31 &   0.26$\pm$0.05 &   d\\
G91.2$+$37.0  & -25.2 &  5.8$\pm$0.3 & 142$\pm$1 &    82$\pm$1 &    60$\pm$1 &   1.09$\pm$0.02 &   0.31 &   0.25$\pm$0.05 &   d
\enddata
\tablenotetext{}{Columns: (1)Known IVMC name; (2) LSR velocity of $^{12}$CO (1-0) emission; (3) Integrated CO line intensity. The angular resolution of the CO observation is 1$^\prime$ for all of the clouds except the Draco clouds (listed in the bottom three rows) for which the resolution is 2.3\arcmin; (4)HI line intensity integrated over $-$100 to 100 \kms (from HI4PI survey data); (5)HI line intensity integrated over $-$20 to 20 \kms (from HI4PI survey data); (6)HI line intensity integrated over $-$100 to $-$20 \kms\ (from HI4PI survey data); (7) HI column density of the IVMC estimated from values given in column 6 (angular resolution $\sim$16.1\arcmin); (8)$A_v$ values obtained from \cite{sch98} data; (9)$A_v$ for the IVMC (see Section~\ref{res}); (10) References from which CO data were obtained for the known IVMCs.  } 
\tablenotetext{}{a)\cite{des90}; b)\cite{rea94}; c)\cite{mag10}; d)\cite{meb85}.}
\tablenotetext{1}{Tentative CO detection \citep[see][]{rea94}.}
\end{deluxetable*}

Despite the high sensitivity of these observations, we did not detect OH emission in any of the individual or summed positions. 
We did detect HI 21-cm emission from both local and intermediate velocities along every sightline we observed. The 3$\sigma$ upper limits of the OH 1667 MHz emission lines are listed in Table~\ref{tab1}. 
\subsection{Derived parameters for candidate IVMCs}
\label{res_cand}
Upper limits on the OH column density are obtained from the 3$\sigma$ line amplitudes given in Table~\ref{tab1} using the equation
\begin{equation}
    N(OH)=2.38 \times 10^{14}\frac{T_{ex}}{T_{ex} - T_{C}} \frac{T_{rms}}{\eta}\; \Delta V \;\;\; \mathrm{cm}^{-2}, 
\end{equation}
where $T_{ex}$ is the excitation temperature of the levels involved, $T_C$ is the background continuum temperature at the cloud location, and $\Delta V$ is the full width at half maximum (FWHM) width of the expected line. The constant 2.38$\times 10^{14}$ comes from the conversion of the integral expression to a product of the peak amplitude and FWHM linewidth. The standard form of this equation is discussed in \citep{lis96}. For this study, we have taken $T_C = 3.1$ K, which we derived from the mean value obtained toward the target clouds from the 408 MHz continuum map \citep{has82}. We first subtracted the CMB temperature from the map value and then scaled to 1667 MHz using a spectral index of 2.65. We then added the CMB temperature to obtain the final value of $T_C$. 
During the observations, we measured the beam efficiency of the Arecibo telescope, $\eta$, by observing bright quasars at 1415 MHz. The average value during the observations was 0.69 and since $\eta$ decreases with frequency, we adopted for this study a value of 0.6 \citep{hei01}. 
For $\Delta V$, we used the smallest intermediate velocity HI line width (see Table~\ref{tab1}). We calculated the final N(OH) values for both $T_{ex}= 5$ and $10$ K. These values are listed in Table~\ref{tab2}.  

As we simultaneously observed the 21-cm line for each target, we obtained the N(HI) directly from our spectra. Given the complexity involved in performing Gaussian decomposition of Galactic hydrogen lines, careful consideration is sometimes required to discern which components should be considered part of the same emission line and ultimately, the same gas population. 
For this study, we selected all Gaussian components within the velocity range $-$90 to $-$20 \kms\ as the HI line components associated with the candidate IVMCs (see Table~\ref{tab1}). We include the LSR velocities provided by \cite{roh16} in Table 2 for comparison with the central velocities of the selected components. 
We obtained the HI column density of each candidate IVMC, N(HI$_{IV}$), using the equation \citep{kul88}
\begin{equation}
    N(HI_{IV})=1.95 \times 10^{18} \sum T_{L}\; \Delta V\;\;\; \mathrm{cm}^{-2}, 
\end{equation}
where $T_{L}$ and $\Delta V$ are the amplitude and FWHM line width of the Gaussian component, respectively, where the summation is over all the components listed in Table~\ref{tab1}. We list the derived N(HI$_{IV}$) of the observed clouds alongside the N(OH) values in Table~\ref{tab2}. Errors are propagated from those obtained during the fitting of the line parameters. 


In the ISM, the dust and gas are thought to be well-mixed, and the correlation between the $A_v$ and the total hydrogen column density is generally understood to be linear for sightlines with low molecular gas fraction i.e. for N(HI) $< 5 \times 10^{20}$ cm$^{-2}$\citep[ see, for example,][] {bou96}. Since the total N(HI) in the directions we observed are less than $5 \times 10^{20}$ cm$^{-2}$, we used this linear relationship to obtain an estimate of $A_v$ associated with our targets in order to ensure that any foreground contribution to the dust is removed. We derived this scaling of the total $A_v$ in the observed sightlines using data from the \cite{sch98} dust maps (6.1\arcmin\ resolution) and the GALFA HI survey data (averaged to the same resolution as the dust maps; see Table~\ref{tab2}). We performed a Gaussian decomposition of the averaged GALFA spectra in the same manner as described for our HI data (see Appendix~\ref{A}) and defined the N(H\small\rm{I}) for low velocity (local) gas from Gaussian components with LSR velocity $>$ $-$20 \kms\  (see Table~\ref{tab5}). The $A_v$ due to this HI column density was then obtained using the relation $A_v=$N(H\small\rm{I})$/$ 2.7$\times 10^{21}$\citep{lis14}. We subtracted this result from the total $A_v$ and assumed that the remaining $A_v$ is associated with intermediate velocity gas having noted visually that there was no discernible contribution from high velocity or positive intermediate velocity gas in the spectra. We henceforth refer to these scaled extinction values as scaled $A_v$, or IV $A_v$. We present these values for each sightline in our study in Table~\ref{tab2}. Uncertainties in the scaled $A_v$ are propagated from the maximum error of the dust map values (16\%) and the error in the N(HI). We also include in Table~\ref{tab2} the HI column density of target IVMCs from the averaged GALFA data. 

We attempted to estimate the $H_2$ column density of the IVMC using the equation 
\begin{equation}
    N(H_2)= \frac{1}{2}\left((2.7 \times 10^{21}\times \mathrm{IV} A_v)- N(HI)\right) \mathrm{cm}^{-2},
    \label{NH2}
\end{equation}
where IV $A_v$ and $N(HI)$ are, respectively, the scaled $A_v$ and HI column density of the intermediate velocity cloud. The N(H$_2$) values estimated this way are strictly upper limits since, in order to obtain the scaled $A_v$ values, we assumed that the foreground contribution to extinction is only due to HI gas, a reasonable assumption for high latitude sight lines. We include in Table~\ref{tab2} the N(H$_2$) values obtained using Eq.~\ref{NH2} for the target sources along with the predicted N(H$_2$) values from \cite{roh16} and upper limits on N(H$_2$) obtained from our OH observations (see Section~\ref{modelcompare}).

\subsection{Observed and derived parameters of known IVMCs}
\label{knownIVMCs}
In addition to the sightlines in this study, we compiled existing data for the known IVMCs detected in CO emission. The LSR velocity of $^{12}$CO emission and the integrated CO line intensity, W$_{CO}$, obtained from the detection papers are listed in Table~\ref{tab3}. The angular resolution of the CO observations are in the range 1\arcmin\ to 2\arcmin. We used the HI4PI data set (angular resolution 16\arcmin) to get the HI properties of the known IVMCs.  The integrated HI intensities for three different velocity ranges -- the total (‘tot’) along the line of sight (from $-$100 to $+$100 \kms), the intermediate velocity (‘IV’; from $-$90 to $-$20 \kms), and the low velocity (‘LV’; from $-$20 to $+$20 \kms) -- are listed in Table~\ref{tab3}. 

\begin{deluxetable*}{lccccc}[htbp]
\tabletypesize{\scriptsize} 
\tablecaption{Physical parameters of known IVMCs from the Diffuse cloud model\label{tab4}}
\tablewidth{300pt}
\tablehead{\colhead{(1)} &\colhead{(2)} & \colhead{(3)} & \colhead{(4)} &  \colhead{(5)} & \colhead{(6)}\\
\colhead{Source} &\colhead{N(H$_2$)} & \colhead{N(H$_2$)} & \colhead{n$_{H}$} &  \colhead{n$_{H}$} & \colhead{N(H$_2$)}\\
&\colhead{(10$^{20}$ cm$^{-2}$)} & \colhead{(10$^{20}$ cm$^{-2}$)} & \colhead{(cm$^{-3}$)} & \colhead{(cm$^{-3}$)} & \colhead{(10$^{20}$ cm$^{-2}$)} }
\startdata
G210.8+63.1 &  1.1$\pm$0.4 & 0.95$\pm$0.08 &  43$\pm$8          &    1140 &    2.5 \\
G135.2+54.5 &  0.7$\pm$0.4 & 0.29$\pm$0.09 &  45$_{-23}^{+8}$   &    1650 &    2.0 \\
G135.5+51.3\tablenotemark{1} &  0.3$\pm$0.3 & 0.26$\pm$0.06 &  21$_{-21}^{+16}$  &     830 &    1.7 \\
G283.8+54.9 &  1.6$\pm$0.6 & 1.01$\pm$0.14 &  48$\pm$10         &     410 &    2.8 \\
G288.4+53.2 &  1.9$\pm$0.6 & 1.35$\pm$0.15 &  74$\pm$14         &     760 &    2.9 \\
G295.0+57.1 &  1.0$\pm$0.4 &  \nodata      & 147$\pm$35         &    1320 &    1.4 \\
G91.3+38.0  &  0.9$\pm$0.3 &  \nodata      & 113$\pm$25         &    5010 &    1.5 \\
G89.8+38.7  &  1.9$\pm$0.5 & 1.66$\pm$0.06 & 128$\pm$18         &    1420 &    2.5 \\
G91.2+37.0  &  2.0$\pm$0.5 & 1.79$\pm$0.08 & 187$\pm$21         &    4280 &    2.5 
\enddata
\tablenotetext{}{Columns: (1)Known IVMC name; (2)Column density of $H_2$ obtained using the scaled $A_v$ and N(HI) of the IVMC; (3)H$_2$ column densities from \cite{roh16}; (4)Total hydrogen density of IVMCs obtained from models using the scaled $A_v$ and N(HI) of IVMC (see Section~\ref{disc}); (5)Upper limit on the total hydrogen density obtained from model using the scaled $A_v$ and W$_{CO}$ (see Section~\ref{disc}); (6)Upper limit on the H$_2$ column density obtained from model using the scaled $A_v$ and W$_{CO}$ (see Section~\ref{disc}).}
\tablenotetext{1}{Tentative CO detection \citep[see][]{rea94}.}
\end{deluxetable*}

We obtained these values and their uncertainties by integrating the HI4PI spectra that can be downloaded from the survey website\footnote{\url{https://www.astro.uni-bonn.de/hisurvey/AllSky gauss/index.php}}. The N(HI$_{IV}$) values were obtained from the integrated HI intensities by dividing by the factor 1.83 $\times$ 10$^{18}$ \citep[see][for the full expression]{kul88}. We further used the foreground N(HI) values to estimate the IV $A_v$ for the known IVMCs as described in Section~\ref{res_cand}. The error in the IV $A_v$ for the known IVMCs is propagated from the uncertainty of integrated HI intensity and that of the dust maps (16\%). The angular resolution of $A_v$ and HI4PI observations differ by a factor of 2.7 and so IV $A_v$ should be used with caution. 

\section{Discussion}
\label{disc}

As mentioned earlier (see Section~\ref{tracers}), the IVMC G211+63 is the only IVMC toward which OH emission was detected to date \citep{smi18}. We compare the properties of this cloud with those of two clouds from our sample. We then present the diffuse cloud model (or photodissociation region (PDR) model), we used and a comparison of its predictions for known IVMCs in Section~\ref{model}. Finally, in \ref{modelcompare}, we apply this model to infer the properties of the candidate IVMCs observed in this study.

\subsection{Comparison with OH emission in IVMC G211+63}
\label{compareclouds}

\begin{figure}[ht!]
\includegraphics[width=1.0\linewidth]{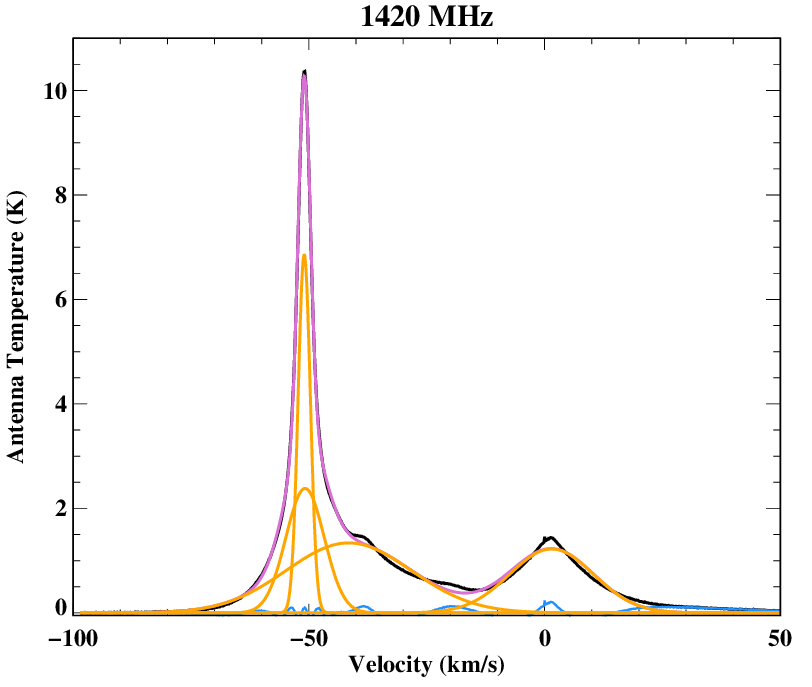}
\caption{The HI 21cm line (black) for candidate IVMC10 with Gaussian components (orange) overplotted. The blue line indicates the residuals from the fit and the magenta line the sum of the Gaussian components.} 
\label{fig3}
\end{figure}

The candidate IVMC10 in our sample (the HI spectrum of which is shown in Fig.~\ref{fig3}) is in many ways similar to the known IVMC G211$+$63, in which OH 18-cm emission was detected in multiple locations \citep{smi18}. In both clouds, the intermediate velocity line dwarfs the local emission near 0 \kms, and the lines are well-fitted by narrow and broad Gaussian components. 
The strongest HI line component of G211$+$63 has amplitude 24.3 K and a FWHM line width of 2.2 \kms (obtained from the GALFA survey data), while the amplitude and line width of IVMC10 are 8 K and 3.97 \kms. The total N(HI) for G211$+$63 and IVMC10 are, respectively, 3.4 $\times$ 10$^{20}$ and 1.62 $\times$ 10$^{20}$ cm$^{-2}$. The total N(HI) value of IVMC10 is also close to the threshold in HI column density (2 $\times$ 10$^{20}$ cm$^{-2}$) noted by \citep{kan11} and beyond which a significant amount of cold HI and possibly molecular gas are expected.  The scaled $A_v$ of G211$+$63 is 0.25 mag, which is, in contrast, 1.9 times higher than that of IVMC10 (whose scaled $A_v$ is 0.13 mag; see Table~\ref{tab2}). The morphological structure of both clouds is compact and well defined in both HI and $A_v$. Despite these broad similarities, no OH was detected at the center position of IVMC10 in our observations despite a very deep integration.

\begin{figure}[ht!]
\includegraphics[width=1.0\linewidth]{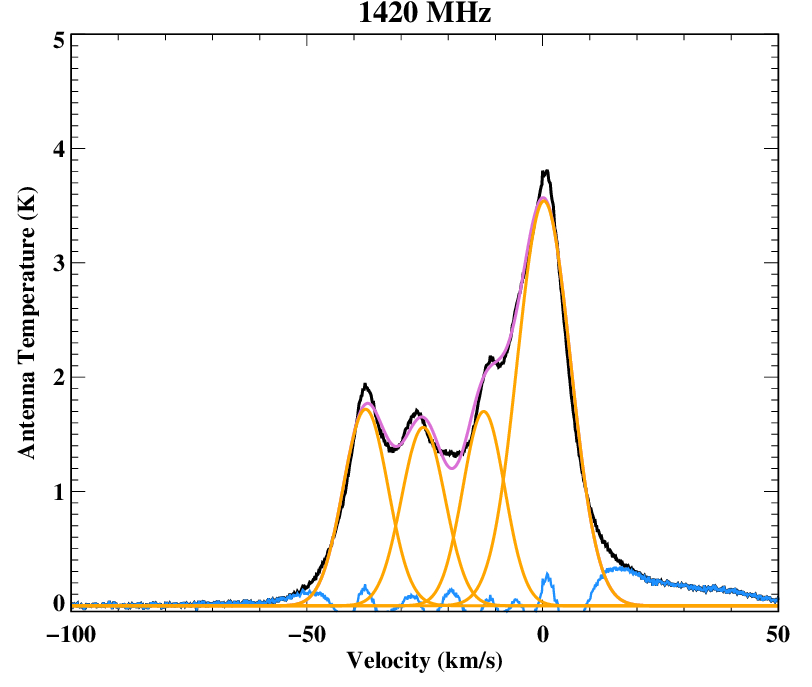}
\caption{The HI 21cm line (black) for candidate IVMC15 with Gaussian components (orange) overplotted. The blue line indicates the residuals from the fit and the magenta line the sum of the Gaussian components.} 
\label{fig4}
\end{figure}

\begin{figure*}[htbp!]
\centering
\includegraphics[width=6in]{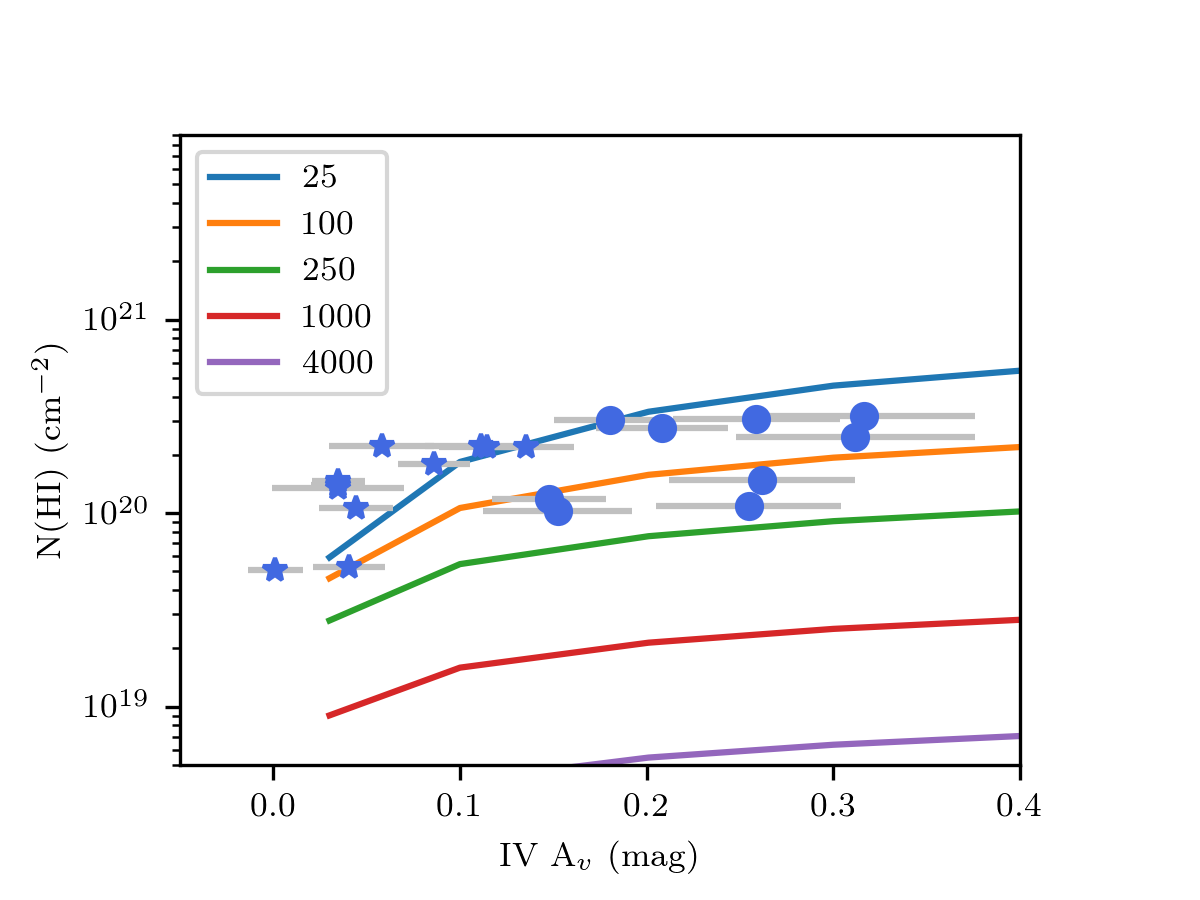}
    \caption{The distribution of the known and candidate IVMCs in the parameter space of $A_v$ and atomic hydrogen column densities. The known IVMCs are represented by circles, while the candidate IVMCs are represented by stars. The N(H\small\rm{I}) and scaled $A_v$ values for IVMCs are derived from observed HI line and extinction images (see Section~\ref{disc}). Gray error bars show the estimated error in scaled $A_v$; the estimation error in N(HI) is within the symbol used for marking the data points. Solid lines represent the model N(HI) vs $A_v$ for different cloud volume densities (marked on the legend in units of cm$^{-3}$).}
    \label{fig5}
\end{figure*}

IVMC15, another candidate in our sample (the HI spectrum of which is shown in Fig.~\ref{fig4}), is an excellent counterexample that showcases the range of HI velocity distributions frequently seen toward IVC sightlines. 
A minimum of four Gaussian components are required to fit the HI spectrum observed toward IVMC15 (see Fig.~\ref{fig4}), and they occupy a velocity range that stretches from the local emission to the expected intermediate velocity line. Also in contrast to the aforementioned clouds, the $A_v$ toward this cloud is 0.09 mag. The N(H\small\rm{I}) value of IVMC15 is 4.5 $\times$ 10$^{19}$ cm$^{-2}$ (see Table~\ref{tab2}), which is 7.6 times lower than that of G211$+$63. 

The sources in our study span a range of N(H\small\rm{I}) and $A_v$.  More than half of the selected sources have HI column density within 60\% of the column density threshold observed in the Galaxy \citep{kan11} where a significant fraction of cold HI and possibly molecular gas are expected. The more extreme cases are analogous to those discussed above.


\subsection{A Diffuse cloud model for IVMCs and comparison with known IVMCs}
\label{model}
In order to constrain the physical parameters of the observed IVMCs, we use the diffuse cloud models of \cite{neu16}.
The models provide in our case the expected N(OH) as a function of volume density of hydrogen nuclei, n$_H$, and visual extinction $A_v$, effectively a measure of the thickness of the cloud. We also make use of the HI, $H_2$ and $^{12}$CO column densities provided by the models for interpreting the observations. 
The models are based on the steady-state photodissociation code described in \cite{hol12}. The ``steady-state" aspect of the models refers in the parlance of \cite{neu16} to the assumption that the time required for the gas species to reach chemical equilibrium is much shorter than the timescales over which the cloud parameters change. For diffuse clouds, the model assumes a FUV radiation field in accordance with that described in \cite{dra78}, and a cosmic ray ionization rate (CRIR) of 2$\times$10$^{-16}$ s~$^{-1}$. For the application of these models to our results, we have assumed the same CRIR, but we assume an FUV field of 0.5 times the Draine value since the gas in question is likely several hundred parsecs away from the plane \citep[see, for e.g.][]{dra78}. The gas temperatures in the model range from 20 K to 110 K depending on the $A_v$ and the density.

We first apply the model results to known IVMCs. Fig.~\ref{fig5} shows the distribution of the IVMCs in the space of scaled $A_v$ and atomic and molecular hydrogen column densities. The PDR model curves -- N(H\small\rm{I}) in solid curves and N(H$_2$) in dashed curves -- are also shown in Fig.~\ref{fig5} for different cloud densities (i.e. total hydrogen nucleus volume density).

The N(H\small\rm{I}) and $A_v$ values are obtained directly from observations, and we apply the PDR models to constrain their corresponding volume densities. The volume densities obtained from the model for the known IVMCs are listed in Table~\ref{tab4}. The known IVMCs appear to cluster between 15$-$250 cm$^{-3}$. We also estimate the H$_2$ column densities for these IVMCs using Eq.~\ref{NH2}, which are also listed in Table~\ref{tab4}. These values are consistent with the model predictions. 

\begin{figure*}[ht!]
\centering
\includegraphics[width=6in]{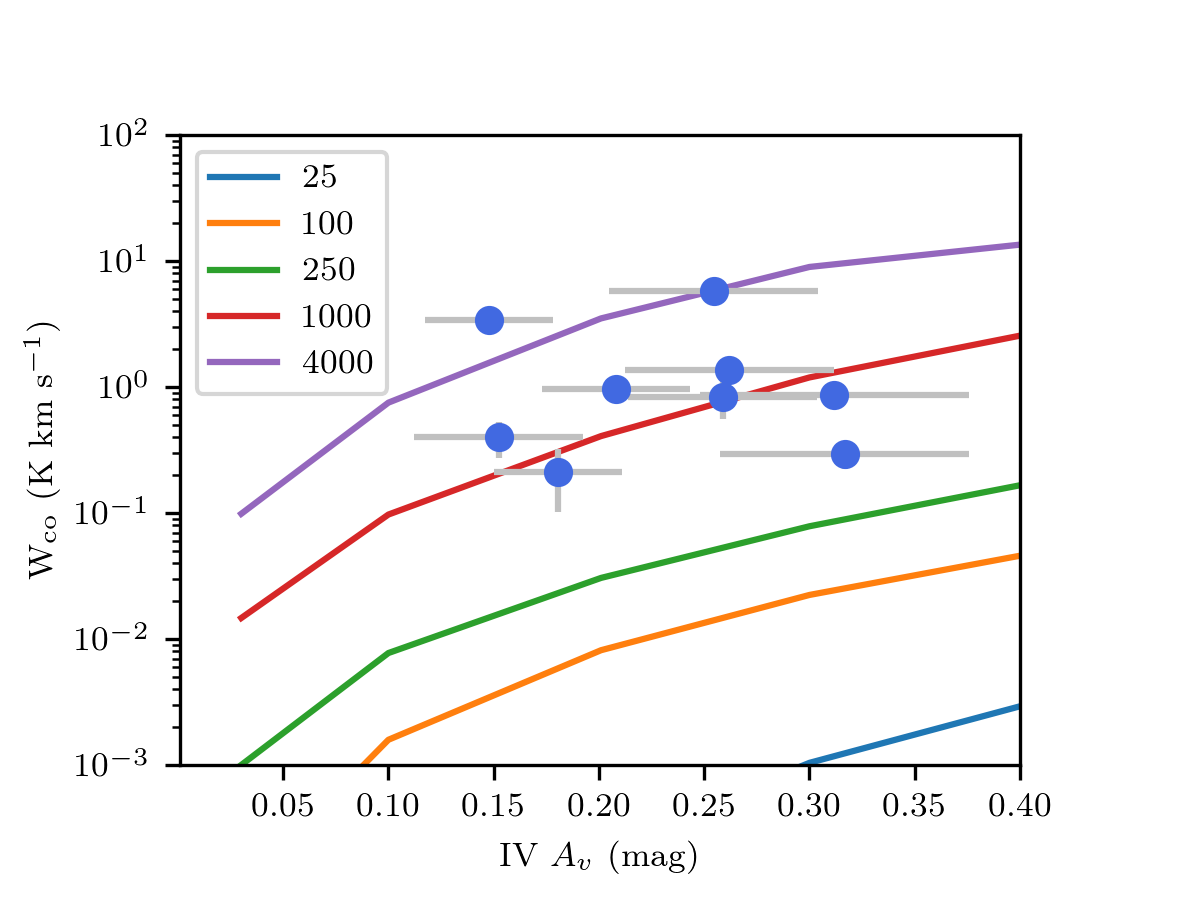}
    \caption{The distribution of the known IVMCs in the parameter space of $A_v$ and W$_{CO}$. The W$_{CO}$ and $A_v$ values are obtained directly from observed CO line emission and extinction images. Gray errors bars show the estimated error in $A_v$; the estimation error in W$_{CO}$ are mostly within the size of the symbol used for marking the data points. Solid lines represent the model $A_v$ vs W$_{CO}$ for different cloud volume densities (marked on the legend in units of cm$^{-3}$). }
    \label{fig6}
\end{figure*}

\begin{figure*}[ht!]
    \centering
    \includegraphics[width=500pt]{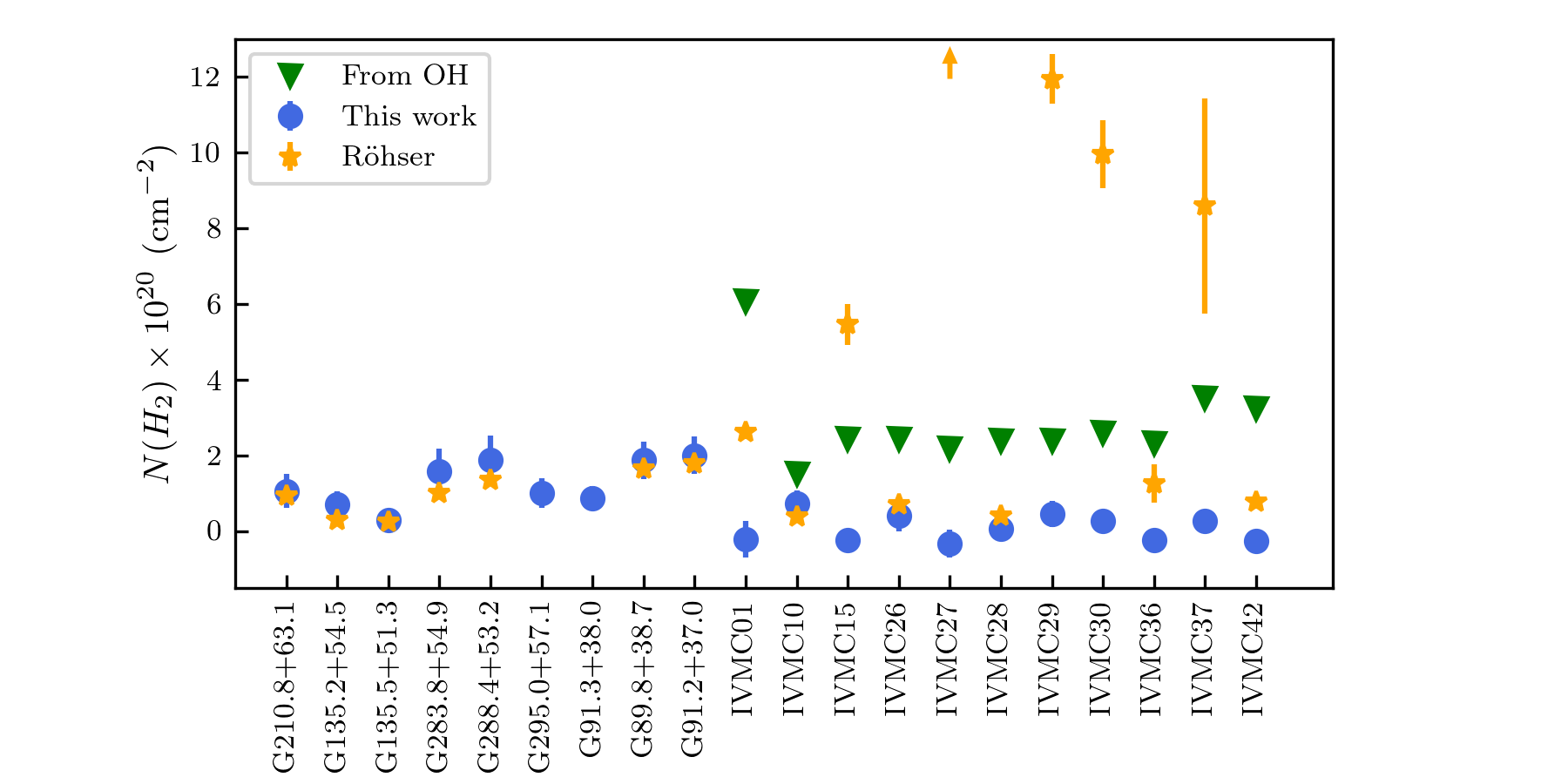}
    \caption{The H$_2$ column densities from this work (marked in blue dots) and from \cite{roh16} (marked in orange stars) for known and candidate IVMCs. For IVMC27, the N(H$_2$) estimate of \cite{roh16} (= 42.1$\pm$14.4 $\times$ 10$^{20}$ cm$^{-2}$)  is outside the y-range of the plot. The N(H$_2$) upper limits obtained from our OH observations toward candidate IVMCs (see Section~\ref{modelcompare}) are shown with green triangles. Error bars that are not visible are smaller than the data points.}
    \label{fig7}
\end{figure*}

In Fig.~\ref{fig7}, we compare the H$_2$ column densities obtained in this work with the estimates for 7 known IVMCs from \cite{roh16} (see Table~\ref{tab4}). Despite the difference in angular resolutions of $A_v$ and HI data for known IVMCs, our estimates compare well with those obtained by \cite{roh16} (see Fig.~\ref{fig7}). 


$^{12}$CO (1-0) emission has been observed toward all the known IVMCs. The integrated CO line emission W$_{CO}$ and the IV $A_v$ values along with the PDR model may also be used to constrain the volume density and N(H$_2$) of the known IVMCs (see Fig.~\ref{fig6}). As seen in Fig.~\ref{fig6}, the known IVMCs cluster around higher $n{_H}$ values compared to their location in the N(H\small\rm{I}) vs IV $A_v$ parameter space (see Fig.~\ref{fig5}). The $n_H$ and N(H$_2$) values obtained from this modeling are listed in Table~\ref{tab4} and listed as upper limits (see below). 

IVMCs may have inhomogeneous molecular structures, and the CO emission is expected to trace the denser areas within these clouds. Our modeling supports this scenario. However, one caveat in this analysis is the difference in angular resolutions of CO (1\arcmin - 2\arcmin), $A_v$ (6\arcmin) and H\small{I} (16\arcmin) observations. The angular extent of denser areas could be smaller compared to the angular resolution of the dust maps. 
Thus, the IV $A_v$ could be higher in those dense regions and hence the $n_H$ and N(H${_2}$) values estimated from Fig.~\ref{fig6} may be considered as upper limits (i.e., the observed points will move toward the right in Fig~\ref{fig6} if IV $A_v$ increases). Nevertheless, the substantially higher $n_H$ values (a factor of $\sim$ 30) compared to those estimated using $A_v$ and N(H\small{I}) is in agreement with the possibility that the candidate IVMCs have small clumps of molecular material embedded in diffuse HI emission.



\begin{figure*}[htbp!]
\includegraphics[width=6in]{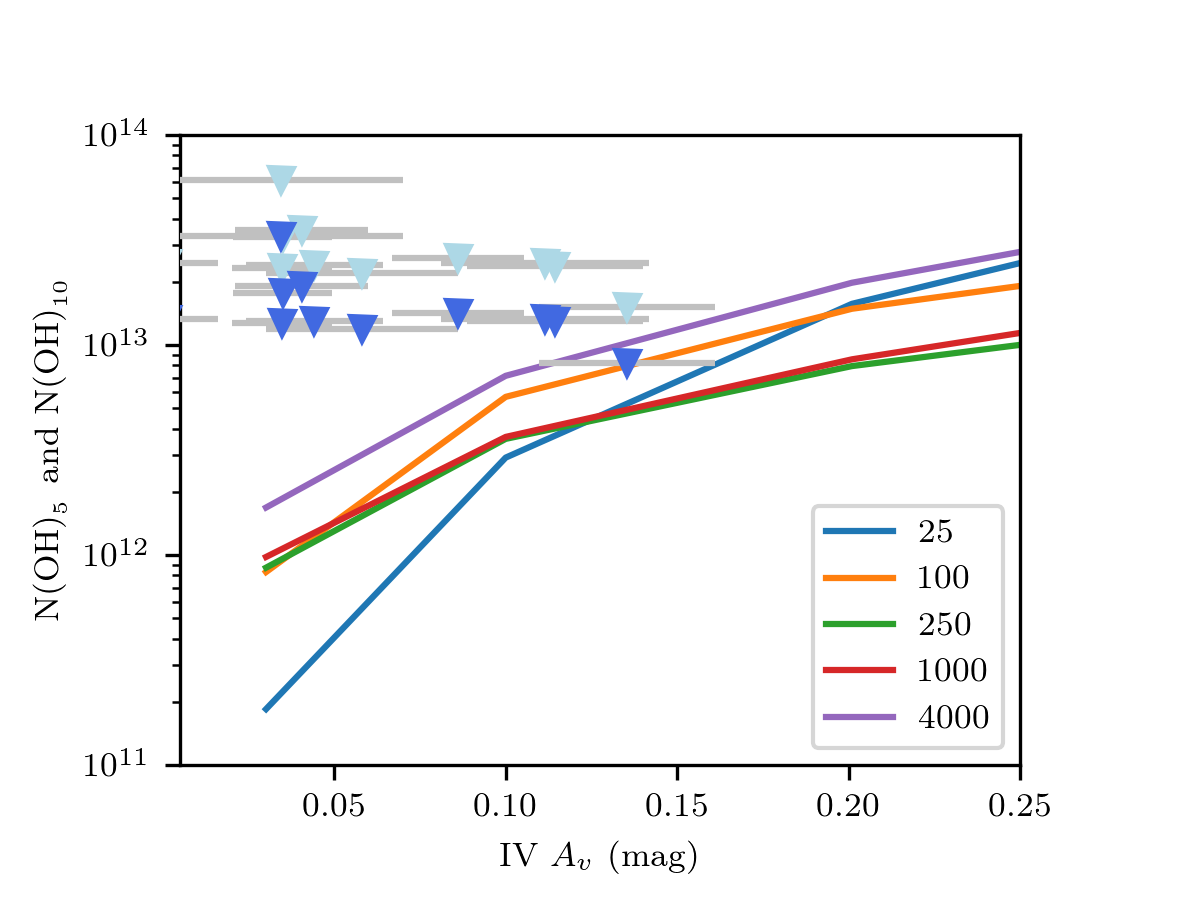}
\centering
    \caption{The 3-$\sigma$ upper limits on OH column densities obtained from the Arecibo data (see Table~\ref{tab2}) are shown in the parameter space N(OH) vs $A_v$ for two excitation temperatures, 5K (light blue) and 10K (dark blue). The uncertainty in $A_v$ is indicated by the gray errors bars. The model N(OH) (solid line) vs $A_v$ are plotted for different cloud volume densities, which are marked on the legend in units of cm$^{-3}$. }
    \label{fig8}
\end{figure*}

\subsection{Application of the Diffuse cloud Model to candidate IVMCs}
\label{modelcompare}
\indent
We now discuss the application of the PDR model to the candidate IVMCs to constrain their physical properties.
Included with the known IVMCs in Fig.~\ref{fig5} are the candidates from our Arecibo study, as indicated by the star symbol. Unlike the case of the known IVMCs, the scaled $A_v$ and N(HI$_{IV}$) for the candidate IVMCs have similar angular resolution, as we obtained GALFA data and smoothed to $\sim$ 6\arcmin\ resolution to obtain N(H\small{I})(see Table~\ref{tab2}).

The candidate IVMCs span a range of scaled $A_v$ 0.01 to 0.1 mag, generally lower than those of the known IVMCs. The N(HI$_{IV}$) values obtained from GALFA data range from 0.5 -- 2.2 $\times 10^{20}$ cm$^{-2}$, significantly overlapping with those obtained for known IVMCs. The volume densities inferred from Fig.~\ref{fig5} are, however, \lsim\ 25 cm$^{-3}$, at least a factor of $\sim$ 1.7 smaller than those obtained for known IVMCs. 

In Fig~\ref{fig7}, we compare the H$_2$ column densities for the candidate IVMCs obtained using Eq.~\ref{NH2} (see Table~\ref{tab2}) with those estimated by \cite{roh16}. Our H$_2$ column densities are significantly lower; the median column density of $H_2$ in the candidate IVMCs is an order of magnitude lower than the estimate of \cite{roh16}. A possible reason for the lower values we obtained is due to the difference in the angular resolution of our observations and the data used by \cite{roh16}. As inferred from the analysis of the known IVMCs, the intermediate velocity clouds are likely highly clumpy on scales of order one arcminute or less. Hence, lacking H\small{I} and $A_v$ data of at least comparable resolution, detecting the molecular gas with single-pointing, targeted observations may be unlikely, as the OH may not be sufficiently widely distributed. While further investigation is required to fully understand the discrepancy between the results in this study and those described in \cite{roh16}, obtaining high resolution ($\sim$ 1\arcmin) H\small{I} data or performing on-the-fly mapping observations of $^{12}$CO emission are of paramount importance for detecting molecular components of IVCs.  

Fig.~\ref{fig8} shows the N(OH) upper limits obtained from the Arecibo observations for two excitation temperatures, 5 K (light blue) and 10 K (dark blue) along with PDR model predictions. Low OH excitation temperatures were recently observed in many Galactic plane molecular clouds  \citep[see, for example,][]{tan20, bus19}. We may expect lower excitation temperature values for the high latitude IVMCs given their diffuse nature and the possibility that OH is widespread but the densities are too low to collisionally populate the levels. 

We converted the N(OH) values to N(H$_2$) upper limits by taking the OH/H$_2$ abundance of $X_{OH} \sim 10^{-7}$ and are listed in Table~\ref{tab2} for the 5 K excitation temperature case. This $X_{OH}$ value is appropriate for the column densities we are considering here \citep{ngu18}.  The estimated N(H$_2$) upper limits are consistent with our inferred lower H{$_2$} column densities in the observed targets (see Fig.~\ref{fig7}). 

\section{Summary}
\label{sum}

We have performed a high sensitivity search of 22 sightlines toward intermediate velocity clouds for OH 18-cm emission using the 305-m telescope at Arecibo Observatory. The 22 targets were selected from the candidate IVMCs compiled by \cite{roh16}. Detailed investigation of these sightlines were made by combining the OH observations with the HI detections in our observations, archival $A_v$ and HI data, as well as utilizing the PDR models of \cite{neu16}. The PDR models were validated using archival $A_v$, HI, and $^{12}$CO data sets toward IVMCs with previous molecular line detections. The summary of our main results is as follows:
\begin{enumerate}
    \item We detected intermediate velocity HI emission in every sightline but no OH emission.
    \item The H$_2$ column densities estimated for known IVMCs compare well with the values obtained by \cite{roh16}, but for target sources they  are more than an order of magnitude lower than estimates by \cite{roh16}. Our OH non-detections are consistent with these lower N(H$_2$) values.
    \item The gas densities inferred using the derived HI column densities and visual extinction for known IVMCs are in the range 15 - 250 cm$^{-3}$. For target sources, similar estimation method provided densities \lsim\ 25 cm$^{-3}$.
    \item For known IVMCs, the gas densities inferred using the CO integrated line intensity and derived visual extinction are at least a factor of $\sim$ 30 larger than those derived using the N(HI)-$A_v$ data set. The substantial difference in density estimates is in agreement with the possibility that the candidate IVMCs have small clumps of molecular material embedded in diffuse HI emission.
\end{enumerate}

We conclude that the candidate IVMCs are unlikely to have widespread molecular gas, and, if present, the structure of the molecular gas will be inhomogeneous with clumps of order of an arcminute or less in angular size.
Higher spatial resolution ($\sim$ 1\arcmin) HI and $A_v$ data are required to better constrain the morphology of the regions within which molecular gas is more likely to be detected in such clouds.
Due to the clumpy nature of the IVMCs, on-the-fly searches for $^{12}$CO emission (rather than pointed observations) with angular resolution of $\sim$ 1\arcmin\ are necessary to detect molecular gas. 

\section{Acknowledgements}
We would like to acknowledge Chris Salter and Tapasi Ghosh for their reading of the manuscript and thoughtful comments. We are also grateful to David Neufeld, who provided the PDR model results that were required to discuss the parameter space occupied by the IVMCs. We are always appreciative of the technical support from the Arecibo Observatory staff; in particular, from Phil Perillat, Arun Venkataraman, and the telescope operators. Finally, we greatly appreciate the anonymous referee who provided extremely thorough and insightful feedback on this manuscript. The Arecibo Observatory is a facility of the NSF operated under cooperative agreement (\#AST-1744119) by the University of Central Florida (UCF) in alliance with Universidad
Ana G. Méndez (UAGM) and Yang Enterprises (YEI), Inc.  

\bibliography{ivoh.bib}

\restartappendixnumbering
\appendix
\section{Gaussian decomposition of the HI line}
\label{A}
This appendix contains the line parameters for the HI 21-cm line from the data obtained in this study and from the averaged (angular resolution 6\arcmin) GALFA spectra toward the observed directions. The complicated structure of Galactic HI requires careful consideration when performing Gaussian decomposition. \cite{ver20} discuss the implications of automated, manual, and a combination of both Gaussian fitting and the difficulty of Gaussian decomposition when working with Galactic HI, and they ultimately determined that semi-automated methods are most likely to yield accurate results. Here we used a semi-automated 
Gaussian fitting routine to decompose the HI spectra.
 We list parameters of all low velocity and intermediate velocity components (indicated by I) for each sightline from our HI data and from the averaged GALFA data in Table~\ref{tab5}. We also include in Fig.~\ref{fig9} the HI spectra for each position we observed.

\startlongtable
\begin{deluxetable*}{lrrrrrrr}
\tablecaption{HI line parameters of observed sightlines\tablenotemark{a}\label{tab5}}
\tablewidth{0pt} 
\tablehead{\colhead{(1)} &\colhead{(2)} &\colhead{(3)}  &\colhead{(4)} &
\colhead{(5)} &\colhead{(6)}  &\colhead{(7)} & \colhead{(8)} \\
\colhead{Source} &\colhead{T$_A$} &\colhead{V$_{LSR}$}  &\colhead{$\Delta$V} &
\colhead{T$_A$} &\colhead{V$_{LSR}$}  &\colhead{$\Delta$V} & \colhead{Population} \\
&\colhead{(K)} &\colhead{(km s$^{-1}$)} &\colhead{(km s$^{-1}$)} &
\colhead{(K)} &\colhead{(km s$^{-1}$)} &\colhead{(km s$^{-1}$)} 
}
\startdata
\hline
\multicolumn{8}{c}{Observed Positions}\\
\hline
IVMC01   &  1.56$\pm$0.03 & -50.75$\pm$0.05 &  6.83$\pm$0.17 &  2.52$\pm$0.11 & -50.91$\pm$0.10 &  2.76$\pm$0.14 & I   \\
        &  1.50$\pm$0.02 & -45.00$\pm$0.32 & 25.49$\pm$0.55 &  2.34$\pm$0.08 & -46.22$\pm$0.54 &  9.65$\pm$0.40 & I   \\
        &  2.90$\pm$0.05 & -11.78$\pm$0.02 &  3.17$\pm$0.06 &  4.14$\pm$0.14 & -11.78$\pm$0.04 &  1.40$\pm$0.06 &   \\
        &  3.74$\pm$0.06 & -5.91$\pm$0.03 &  3.49$\pm$0.07 &  5.43$\pm$0.17 & -5.91$\pm$0.06 &  1.58$\pm$0.07 &   \\
        &  8.83$\pm$0.09 & -1.91$\pm$0.01 &  3.02$\pm$0.03 & 12.13$\pm$0.28 & -1.92$\pm$0.03 &  1.27$\pm$0.03 &   \\
        &  6.77$\pm$0.05 &  2.47$\pm$0.03 &  6.38$\pm$0.08 &  9.68$\pm$0.14 &  2.47$\pm$0.07 &  2.78$\pm$0.07 &   \\
        &  3.06$\pm$0.06 & -8.06$\pm$0.21 & 31.01$\pm$0.46 &  3.71$\pm$0.13 & -9.55$\pm$0.51 & 14.10$\pm$0.44 &   \\
IVMC01N  &  5.36$\pm$0.03 & -51.78$\pm$0.01 &  5.80$\pm$0.04 & \nodata & \nodata & \nodata   &   I   \\
        &  2.34$\pm$0.02 & -47.50$\pm$0.12 & 22.92$\pm$0.25 & \nodata & \nodata & \nodata   &   I   \\
        &  4.16$\pm$0.04 & -11.47$\pm$0.01 &  2.73$\pm$0.03 & \nodata & \nodata & \nodata   &     \\
        &  4.24$\pm$0.11 & -5.68$\pm$0.03 &  3.57$\pm$0.06 & \nodata & \nodata & \nodata   &     \\
        &  7.28$\pm$0.18 & -1.76$\pm$0.01 &  2.76$\pm$0.04 & \nodata & \nodata & \nodata   &     \\
        &  7.35$\pm$0.08 &  1.13$\pm$0.08 &  7.84$\pm$0.13 & \nodata & \nodata & \nodata   &     \\
        &  4.61$\pm$0.04 & -8.72$\pm$0.09 & 30.34$\pm$0.19 & \nodata & \nodata & \nodata   &     \\
IVMC01E &  4.35$\pm$0.05 & -49.40$\pm$0.02 &  3.20$\pm$0.05 & \nodata & \nodata & \nodata   &   I   \\
        &  2.06$\pm$0.03 & -48.65$\pm$0.14 & 20.43$\pm$0.35 & \nodata & \nodata & \nodata   &   I   \\
        &  5.45$\pm$0.06 & -11.31$\pm$0.02 &  3.77$\pm$0.05 & \nodata & \nodata & \nodata   &     \\
        &  7.73$\pm$0.07 & -5.93$\pm$0.01 &  2.93$\pm$0.04 & \nodata & \nodata & \nodata   &     \\
        &  9.20$\pm$0.18 & -1.91$\pm$0.01 &  3.09$\pm$0.05 & \nodata & \nodata & \nodata   &     \\
        &  8.19$\pm$0.06 &  2.13$\pm$0.06 &  7.23$\pm$0.11 & \nodata & \nodata & \nodata   &     \\
        &  3.73$\pm$0.05 & -10.12$\pm$0.17 & 34.87$\pm$0.35 & \nodata & \nodata & \nodata   &     \\
IVMC01W &  3.89$\pm$0.07 & -51.02$\pm$0.03 &  6.70$\pm$0.12 & \nodata & \nodata & \nodata   &   I   \\
        &  3.03$\pm$0.06 & -46.70$\pm$0.26 & 22.87$\pm$0.52 & \nodata & \nodata & \nodata   &   I   \\
        &  2.85$\pm$0.03 & -19.90$\pm$0.24 & 18.17$\pm$0.78 & \nodata & \nodata & \nodata   &   I   \\
        &  2.00$\pm$0.00 & -12.14$\pm$0.04 &  2.90$\pm$0.12 & \nodata & \nodata & \nodata   &     \\
        &  7.48$\pm$0.11 & -1.53$\pm$0.01 &  2.87$\pm$0.04 & \nodata & \nodata & \nodata   &     \\
        &  5.14$\pm$0.10 &  3.41$\pm$0.02 &  2.92$\pm$0.06 & \nodata & \nodata & \nodata   &     \\
        &  8.88$\pm$0.09 & -1.01$\pm$0.10 & 15.54$\pm$0.19 & \nodata & \nodata & \nodata   &     \\
IVMC10  &  8.01$\pm$0.05 & -50.98$\pm$0.01 &  3.97$\pm$0.03 & 10.68$\pm$0.13 & -51.02$\pm$0.02 &  1.71$\pm$0.03 & I   \\
        &  1.92$\pm$0.02 & -45.55$\pm$0.13 & 26.73$\pm$0.30 &  2.61$\pm$0.07 & -45.45$\pm$0.28 & 11.27$\pm$0.27 & I   \\
        &  1.15$\pm$0.02 &  0.62$\pm$0.19 & 25.03$\pm$0.46 &  1.52$\pm$0.05 &  0.41$\pm$0.39 &  9.54$\pm$0.39 &   \\
IVMC15  &  1.40$\pm$0.02 & -37.91$\pm$0.07 &  8.55$\pm$0.13 &  2.78$\pm$0.05 & -38.06$\pm$0.07 &  2.80$\pm$0.07 & I   \\
        &  1.02$\pm$0.02 & -26.63$\pm$0.10 & 10.89$\pm$0.35 &  0.90$\pm$0.08 & -27.68$\pm$0.26 &  3.72$\pm$0.42 & I   \\
        &  1.23$\pm$0.03 & -10.14$\pm$0.17 & 11.82$\pm$0.42 &  0.99$\pm$0.11 & -9.99$\pm$0.40 &  3.88$\pm$0.51 &   \\
        &  2.84$\pm$0.03 &  0.75$\pm$0.07 &  9.91$\pm$0.11 &  4.05$\pm$0.06 &  0.63$\pm$0.11 &  4.16$\pm$0.09 &   \\
        &  0.80$\pm$0.03 & -9.59$\pm$0.30 & 52.96$\pm$0.67 &  1.81$\pm$0.11 & -15.45$\pm$0.47 & 17.46$\pm$0.39 &   \\
IVMC26  &  4.04$\pm$0.13 & -35.55$\pm$0.05 &  7.30$\pm$0.15 &  2.83$\pm$0.12 & -36.47$\pm$0.07 &  2.14$\pm$0.10 & I   \\
        &  4.03$\pm$0.07 & -25.37$\pm$0.04 &  5.13$\pm$0.10 &  3.79$\pm$0.10 & -24.86$\pm$0.04 &  1.35$\pm$0.04 & I   \\
        &  2.00$\pm$0.00 & -28.88$\pm$0.50 & 15.21$\pm$0.66 &  5.62$\pm$0.06 & -29.62$\pm$0.14 &  6.58$\pm$0.08 & I   \\
        &  4.72$\pm$0.14 & -2.98$\pm$0.05 &  9.58$\pm$0.20 &  2.69$\pm$0.11 & -4.49$\pm$0.06 &  1.71$\pm$0.08 &   \\
        &  1.88$\pm$0.15 & -1.50$\pm$0.29 & 22.14$\pm$0.79 &  7.09$\pm$0.10 & -2.37$\pm$0.05 &  5.86$\pm$0.06 &   \\
IVMC27  &  1.04$\pm$0.02 & -40.97$\pm$0.78 & 29.44$\pm$1.27 &  0.15$\pm$0.09 & -56.19$\pm$2.94 &  6.29$\pm$4.13 & I   \\
        &  3.23$\pm$0.05 & -32.70$\pm$0.03 &  4.93$\pm$0.09 &  4.57$\pm$0.10 & -32.82$\pm$0.05 &  2.02$\pm$0.06 & I   \\
        &  1.00$\pm$0.00 & -20.42$\pm$0.41 & 14.37$\pm$0.83 &  1.92$\pm$0.16 & -21.47$\pm$1.99 & 19.87$\pm$0.86 & I   \\
        &  2.82$\pm$0.03 & -4.49$\pm$0.06 &  8.98$\pm$0.14 &  5.06$\pm$1.33 & -4.51$\pm$0.11 &  4.42$\pm$0.37 &   \\
        &  2.50$\pm$0.00 & -4.97$\pm$0.27 & 22.39$\pm$0.43 &  1.79$\pm$1.21 & -4.88$\pm$0.77 &  7.84$\pm$2.29 &   \\
IVMC28  &  1.91$\pm$0.02 & -44.48$\pm$0.02 &  5.02$\pm$0.07 &  2.66$\pm$0.08 & -44.48$\pm$0.07 &  2.35$\pm$0.08 & I   \\
        &  1.52$\pm$0.02 & -46.41$\pm$0.24 & 23.59$\pm$0.36 &  1.85$\pm$0.13 & -48.20$\pm$0.62 &  9.08$\pm$0.35 & I   \\
        &  5.18$\pm$0.02 & -18.54$\pm$0.01 &  3.95$\pm$0.02 &  7.53$\pm$0.08 & -18.62$\pm$0.02 &  1.73$\pm$0.02 &   \\
        &  1.50$\pm$0.02 & -20.00$\pm$0.00 & 21.25$\pm$0.83 &  2.27$\pm$0.05 & -17.05$\pm$0.42 & 14.60$\pm$1.30 &   \\
        &  0.70$\pm$0.04 & -4.13$\pm$0.24 & 18.14$\pm$0.38 & -0.28$\pm$0.12 &  8.01$\pm$1.21 &  5.74$\pm$1.97 &   \\
IVMC28N &  1.50$\pm$0.00 & -45.80$\pm$0.04 &  6.69$\pm$0.11 & \nodata & \nodata & \nodata   &   I   \\
        &  1.05$\pm$0.02 & -50.99$\pm$0.26 & 26.98$\pm$0.42 & \nodata & \nodata & \nodata   &   I   \\
        &  4.97$\pm$0.02 & -18.13$\pm$0.01 &  3.95$\pm$0.02 & \nodata & \nodata & \nodata   &     \\
        &  1.52$\pm$0.01 & -17.21$\pm$0.22 & 36.80$\pm$0.41 & \nodata & \nodata & \nodata   &     \\
IVMC28S &  1.50$\pm$0.00 & -44.45$\pm$0.09 &  6.87$\pm$0.22 & \nodata & \nodata & \nodata   &   I   \\
        &  0.78$\pm$0.04 & -51.77$\pm$0.22 & 10.79$\pm$0.47 & \nodata & \nodata & \nodata   &   I   \\
        &  1.27$\pm$0.02 & -28.19$\pm$0.71 & 49.07$\pm$0.69 & \nodata & \nodata & \nodata   &   I   \\
        &  6.82$\pm$0.04 & -19.19$\pm$0.01 &  4.20$\pm$0.03 & \nodata & \nodata & \nodata   &     \\
        &  0.68$\pm$0.04 & -10.00$\pm$0.00 & 19.70$\pm$1.15 & \nodata & \nodata & \nodata   &     \\
IVMC28E &  1.25$\pm$0.04 & -46.07$\pm$0.06 &  9.01$\pm$0.22 & \nodata & \nodata & \nodata   &   I   \\
        &  0.97$\pm$0.04 & -50.74$\pm$0.36 & 21.03$\pm$0.39 & \nodata & \nodata & \nodata   &   I   \\
        &  5.15$\pm$0.02 & -18.72$\pm$0.01 &  4.16$\pm$0.02 & \nodata & \nodata & \nodata   &     \\
        &  1.60$\pm$0.01 & -19.25$\pm$0.20 & 36.32$\pm$0.38 & \nodata & \nodata & \nodata   &     \\
        &  0.16$\pm$0.02 & -2.78$\pm$0.23 &  4.46$\pm$0.60 & \nodata & \nodata & \nodata   &     \\
IVMC28W &  2.01$\pm$0.04 & -44.09$\pm$0.05 &  5.92$\pm$0.15 & \nodata & \nodata & \nodata   &   I   \\
        &  1.33$\pm$0.03 & -41.82$\pm$0.48 & 31.69$\pm$1.05 & \nodata & \nodata & \nodata   &   I   \\
        &  0.54$\pm$0.05 & -27.09$\pm$0.24 &  7.31$\pm$0.82 & \nodata & \nodata & \nodata   &   I   \\
        &  4.80$\pm$0.04 & -18.06$\pm$0.02 &  5.55$\pm$0.06 & \nodata & \nodata & \nodata   &     \\
        &  1.39$\pm$0.02 & -6.95$\pm$0.17 & 18.00$\pm$0.00 & \nodata & \nodata & \nodata   &     \\
IVMC29  &  2.49$\pm$0.01 & -30.45$\pm$0.09 & 27.24$\pm$0.18 &  3.28$\pm$0.03 & -29.93$\pm$0.15 & 11.85$\pm$0.14 & I   \\
        &  4.45$\pm$0.03 & -22.71$\pm$0.01 &  3.95$\pm$0.03 &  5.53$\pm$0.07 & -23.09$\pm$0.02 &  1.60$\pm$0.03 & I   \\
        &  6.01$\pm$0.04 & -0.91$\pm$0.01 &  3.16$\pm$0.02 &  9.57$\pm$0.10 & -0.98$\pm$0.01 &  1.25$\pm$0.02 &   \\
        &  2.05$\pm$0.03 & -2.75$\pm$0.07 & 13.24$\pm$0.17 &  2.88$\pm$0.07 & -2.51$\pm$0.11 &  5.18$\pm$0.12 &   \\
IVMC30  &  4.14$\pm$0.05 & -29.59$\pm$0.03 &  4.56$\pm$0.07 &  5.07$\pm$0.11 & -29.52$\pm$0.05 &  2.11$\pm$0.06 & I   \\
        &  1.27$\pm$0.02 & -18.95$\pm$0.38 & 43.47$\pm$0.70 &  1.64$\pm$0.04 & -19.88$\pm$0.67 & 17.46$\pm$0.53 & I   \\
        &  4.97$\pm$0.04 & -0.29$\pm$0.03 &  7.53$\pm$0.08 &  6.16$\pm$0.09 & -0.35$\pm$0.05 &  3.23$\pm$0.06 &   \\
IVMC36  &  2.70$\pm$0.00 & -29.23$\pm$0.05 & 14.72$\pm$0.11 &  2.66$\pm$0.07 & -29.34$\pm$0.13 &  8.56$\pm$0.17 & I   \\
        &  1.77$\pm$0.03 & -27.79$\pm$0.04 &  3.49$\pm$0.08 &  3.51$\pm$0.08 & -28.02$\pm$0.04 &  2.14$\pm$0.06 & I   \\
        &  5.04$\pm$0.04 &  0.61$\pm$0.02 &  5.20$\pm$0.05 &  5.70$\pm$0.12 &  0.99$\pm$0.03 &  2.31$\pm$0.05 &   \\
        &  1.81$\pm$0.04 & -0.75$\pm$0.10 & 19.87$\pm$0.33 &  2.81$\pm$0.12 & -0.09$\pm$0.13 &  6.88$\pm$0.18 &   \\
IVMC37  &  3.00$\pm$0.00 & -30.26$\pm$0.05 &  7.39$\pm$0.13 &  3.69$\pm$0.11 & -30.33$\pm$0.10 &  3.13$\pm$0.12 & I   \\
        &  7.42$\pm$0.05 &  0.91$\pm$0.02 &  5.88$\pm$0.05 &  9.24$\pm$0.12 &  0.91$\pm$0.04 &  2.55$\pm$0.04 &   \\
        &  1.01$\pm$0.02 & -15.74$\pm$0.52 & 46.85$\pm$0.93 &  1.27$\pm$0.05 & -18.13$\pm$1.06 & 18.24$\pm$0.77 &   \\
IVMC42  &  4.41$\pm$0.03 & -28.98$\pm$0.02 &  6.02$\pm$0.05 &  5.87$\pm$0.09 & -29.01$\pm$0.03 &  2.14$\pm$0.04 & I   \\
        &  0.85$\pm$0.01 & -19.95$\pm$0.45 & 43.37$\pm$0.85 &  2.08$\pm$0.08 & -27.93$\pm$0.18 &  9.24$\pm$0.25 & I   \\
        &  2.00$\pm$0.00 & -2.64$\pm$0.05 &  3.48$\pm$0.12 &  2.72$\pm$0.13 & -2.63$\pm$0.05 &  1.26$\pm$0.07 &   \\
        &  4.74$\pm$0.04 &  0.40$\pm$0.03 &  8.26$\pm$0.08 &  7.09$\pm$0.10 & -0.10$\pm$0.06 &  4.09$\pm$0.04 &   \\
IVMC42N &  3.83$\pm$0.04 & -28.91$\pm$0.02 &  5.71$\pm$0.07 & \nodata & \nodata & \nodata   &   I   \\
        &  1.22$\pm$0.03 & -28.22$\pm$0.14 & 24.70$\pm$0.49 & \nodata & \nodata & \nodata   &   I   \\
        &  3.00$\pm$0.00 & -1.81$\pm$0.03 &  4.32$\pm$0.08 & \nodata & \nodata & \nodata   &     \\
        &  4.04$\pm$0.03 &  0.78$\pm$0.03 & 10.11$\pm$0.08 & \nodata & \nodata & \nodata   &     \\
IVMC42S &  3.16$\pm$0.04 & -28.67$\pm$0.02 &  3.48$\pm$0.05 & \nodata & \nodata & \nodata   &   I   \\
        &  1.41$\pm$0.03 & -28.72$\pm$0.09 & 17.84$\pm$0.28 & \nodata & \nodata & \nodata   &   I   \\
        &  5.10$\pm$0.03 & -0.35$\pm$0.02 &  8.36$\pm$0.05 & \nodata & \nodata & \nodata   &     \\
        &  0.50$\pm$0.02 &  0.00$\pm$0.00 & 37.39$\pm$1.44 & \nodata & \nodata & \nodata   &     \\
IVMC42E &  4.69$\pm$0.04 & -28.74$\pm$0.02 &  4.87$\pm$0.06 & \nodata & \nodata & \nodata   &   I   \\
        &  1.21$\pm$0.03 & -26.89$\pm$0.17 & 22.41$\pm$0.51 & \nodata & \nodata & \nodata   &   I   \\
        & -0.71$\pm$0.05 & -4.69$\pm$0.11 &  3.52$\pm$0.26 & \nodata & \nodata & \nodata   &     \\
        &  5.78$\pm$0.03 &  0.00$\pm$0.00 &  9.51$\pm$0.06 & \nodata & \nodata & \nodata   &     \\
IVMC42W &  2.74$\pm$0.04 & -28.91$\pm$0.02 &  4.21$\pm$0.07 & \nodata & \nodata & \nodata   &   I   \\
        &  1.60$\pm$0.03 & -28.87$\pm$0.13 & 18.53$\pm$0.36 & \nodata & \nodata & \nodata   &   I   \\
        &  5.41$\pm$0.04 & -0.94$\pm$0.02 &  8.10$\pm$0.06 & \nodata & \nodata & \nodata   &     \\
        &  0.72$\pm$0.03 &  1.38$\pm$0.40 & 26.97$\pm$0.97 & \nodata & \nodata & \nodata   &     \\
\hline
\multicolumn{8}{c}{Summed Positions}\\
\hline
IVMC01T &  2.45$\pm$0.02 & -50.75$\pm$0.03 &  7.48$\pm$0.09 & \nodata & \nodata & \nodata   &   I   \\
        &  1.19$\pm$0.01 & -39.38$\pm$0.37 & 35.83$\pm$0.66 & \nodata & \nodata & \nodata   &   I   \\
        &  4.16$\pm$0.07 & -1.34$\pm$0.08 &  5.53$\pm$0.13 & \nodata & \nodata & \nodata   &     \\
        &  4.37$\pm$0.03 & -4.97$\pm$0.08 & 22.46$\pm$0.14 & \nodata & \nodata & \nodata   &     \\
        &  2.35$\pm$0.09 &  3.54$\pm$0.13 &  5.01$\pm$0.17 & \nodata & \nodata & \nodata   &     \\
IVMC28T &  1.50$\pm$0.00 & -44.15$\pm$0.05 &  7.24$\pm$0.13 & \nodata & \nodata & \nodata   &   I   \\
        &  1.22$\pm$0.02 & -44.90$\pm$0.35 & 28.50$\pm$0.58 & \nodata & \nodata & \nodata   &   I   \\
        &  3.54$\pm$0.02 & -17.48$\pm$0.02 &  6.06$\pm$0.05 & \nodata & \nodata & \nodata   &     \\
        &  1.56$\pm$0.02 & -12.99$\pm$0.30 & 29.71$\pm$0.43 & \nodata & \nodata & \nodata   &     \\
IVMC42T &  3.39$\pm$0.03 & -28.86$\pm$0.02 &  4.36$\pm$0.05 & \nodata & \nodata & \nodata   &   I   \\
        &  1.58$\pm$0.03 & -28.77$\pm$0.11 & 18.10$\pm$0.32 & \nodata & \nodata & \nodata   &   I   \\
        &  5.33$\pm$0.03 & -0.51$\pm$0.02 &  8.19$\pm$0.05 & \nodata & \nodata & \nodata   &     \\
        &  0.62$\pm$0.03 &  0.47$\pm$0.40 & 28.18$\pm$1.12 & \nodata & \nodata & \nodata   &     \\
\enddata 
\tablenotetext{}Columns: (1)Target name; (2), (3) \& (4) Line temperatures, LSR velocities and FWHM line widths, respectively, of Gaussian components obtained from modeling our data (angular resolution $\sim$ 3\arcmin); (5), (6), \& (7) Line temperatures, LSR velocities and FWHM line widths, respectively, of Gaussian components obtained from modeling averaged GALFA spectrum (angular resolution $\sim$ 6\arcmin); (8)We identify the components we consider indicative of an intermediate-velocity gas population with an I.
\tablenotetext{a}For components associated with IVMCs, the parameters of Gaussian fits to our data are also given in Table~\ref{tab1} but are repeated here for completeness. Parameters with no errors are those that were fixed while fitting Gaussian components to the spectrum.
\end{deluxetable*}

\begin{figure*}
    \centering
    \includegraphics[width=\textwidth]{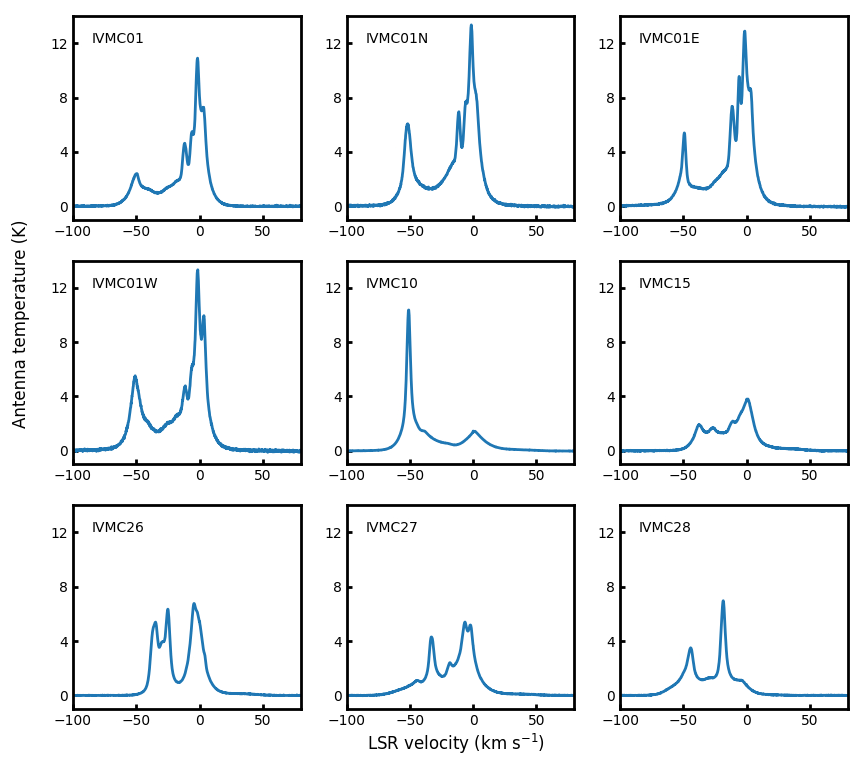}
    \label{a1}
\end{figure*}
\begin{figure*}
    \centering
    \includegraphics[width=\textwidth]{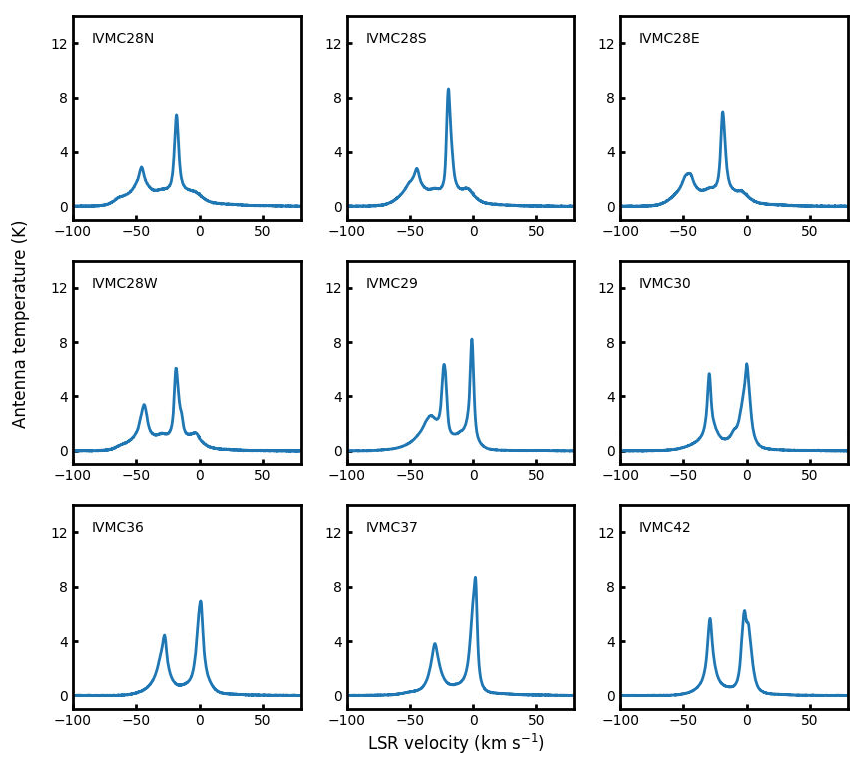}
    \label{a2}
\end{figure*}
\begin{figure*}
    \centering
    \includegraphics[width=\textwidth]{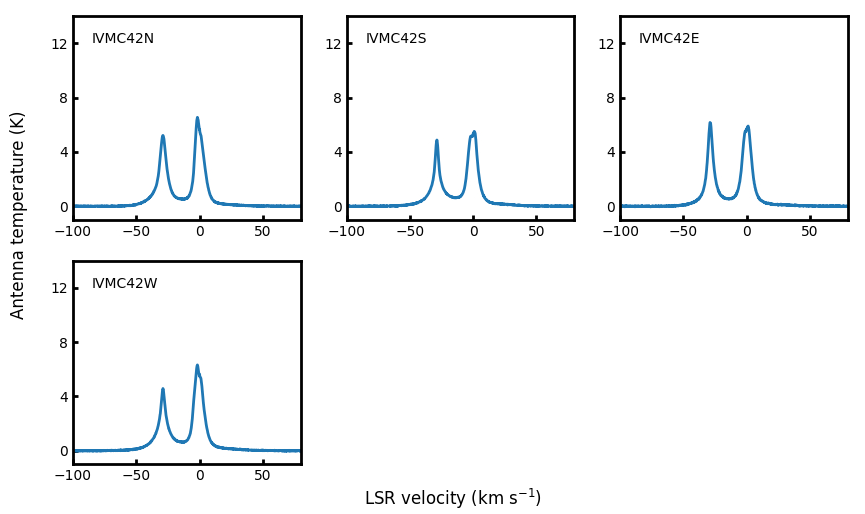}
    \caption{The Arecibo HI 21-cm line spectra of all sightlines. Both local ($\sim$0 \kms) and IV emission are apparent.}
    \label{fig9}
\end{figure*}

\end{document}